\documentclass[iop]{emulateapj}
\usepackage{longtable,subfigure,float}

\usepackage[bookmarks=false]{hyperref}

\def\micron{{\mbox{$\mu{\rm m}$}}}
\def\arcsec{{\mbox{$^{\prime \prime}$}}}
\def\arcmin{{\mbox{$^{\prime}$}}}
\def\degree{{\mbox{$^{\circ}$}}}

\def \nustar {{\it NuSTAR\ }}

\def \swiftbat {{\it Swift} BAT\ }
\def \Chr {{\it Chandra\ }}
\def \chandra {{\it Chandra\ }}
\def \XMM{{\it XMM-Newton\ }}

\def \nustarsh {{\it NuSTAR}}

\def \Chrsh {{\it Chandra}}

\def \XMMsh{{\it XMM-Newton}}

\def \xspec {{\tt Xspec\ }}
\def \mytorus {{\tt MYtorus\ }}
\def \bntorus {{\tt BNtorus\ }}
\def \pexrav {{\tt Pexrav\ }}
\def \hst{{\it HST\ }}
\def \spitzer{{\it Spitzer\ }}

\def \bepposax{{\it BeppoSAX\ }}
\def \bepposaxsh{{\it BeppoSAX}}
\def \suzaku{{\it Suzaku\ }}
\def \suzakush{{\it Suzaku}}

\newcommand{\oiii}{[O\,{\sc iii}] }
\newcommand{\oiiish}{[O\,{\sc iii}]}

\newcommand{\nev}{[Ne\,{\sc v}] }
\newcommand{\oiv}{[O\,{\sc iv}] }

\def\arcsec{{\mbox{$^{\prime \prime}$}}}
\def\cm{{\rm\thinspace cm}}

\def\erg{{\rm\thinspace erg}}

\def\km{{\rm\thinspace km}}

\def\Msun{\hbox{$\rm\thinspace M_{\odot}$}}

\def\s{{\rm\thinspace s}}

\def\ergps{\hbox{$\erg\s^{-1}\,$}}

\def\pcmsq{\hbox{$\cm^{-2}\,$}}

\def\cps{\hbox{cts$\,\s^{-1}$}}
\def\micron{{\mbox{$\mu{\rm m}$}}}
\def\arcsec{{\mbox{$^{\prime \prime}$}}}
\def\arcmin{{\mbox{$^{\prime}$}}}
\def\degree{{\mbox{$^{\circ}$}}}

\def\kmpssh{\hbox{$\km\s^{-1}$}}
\def\ergpssh{\hbox{$\erg\s^{-1}$}}
\def\cmsqsh{\hbox{$\cm^2$}}
\def\pcmsqsh{\hbox{$\cm^{-2}$}}

\begin{document}
\title{Broadband Observations of the Compton-thick Nucleus of NGC 3393}

\author{Michael J. Koss\altaffilmark{1,2,20},  C.\ Romero-Ca\~nizales\altaffilmark{3,4},  L. Baronchelli\altaffilmark{1}, S. H. Teng\altaffilmark{5,6,21}, M. Balokovi{\' c}\altaffilmark{7}, S. Puccetti\altaffilmark{8,9}, F.E. Bauer\altaffilmark{4,10}, P. Ar\'evalo\altaffilmark{11}, R. Assef\altaffilmark{12}, D.R. Ballantyne\altaffilmark{13},  W. N. Brandt\altaffilmark{14}, M. Brightman\altaffilmark{7}, A. Comastri\altaffilmark{15}, P. Gandhi\altaffilmark{16,17}, F. A. Harrison\altaffilmark{7}, B. Luo\altaffilmark{13}, K. Schawinski\altaffilmark{1}, D. Stern\altaffilmark{18}, and E. Treister\altaffilmark{19}}
\email{mkoss@phys.ethz.ch}

\altaffiltext{1}{Institute for Astronomy, Department of Physics, ETH Zurich, Wolfgang-Pauli-Strasse 27, CH-8093 Zurich, Switzerland; mkoss@phys.ethz.ch}
\altaffiltext{2}{Institute for Astronomy, University of Hawaii, 2680 Woodlawn Drive, Honolulu, HI 96822, USA}
\altaffiltext{3}{Millennium Institute of Astrophysics, Vicu\~na Mackenna 4860, 7820436, Macul, Santiago, Chile}
\altaffiltext{4}{Instituto de Astrof\'{\i}sica, Facultad de F\'{\i}sica, Pontificia Universidad Cat\'olica de Chile, Casilla 306, Santiago 22, Chile}
\altaffiltext{5}{Observational Cosmology Laboratory, NASA Goddard Space Flight Center, Greenbeltt, MD 20771, USA}
\altaffiltext{6}{CRESST, Department of Astronomy, University of Maryland, College Park, MD 20742}
\altaffiltext{7}{Cahill Center for Astronomy and Astrophysics, California Institute of Technology, Pasadena, CA 91125, USA}
\altaffiltext{8}{INAFÐOsservatorio Astronomico di Roma, via Frascati 33, 00040 Monte Porzio Catone (RM), Italy}
\altaffiltext{9}{ASDC--ASI, Via del Politecnico, 00133 Roma, Italy}
\altaffiltext{10}{Space Science Institute, 4750 Walnut Street, Suite 205, Boulder, CO 80301, USA}
\altaffiltext{11}{Instituto de F\'isica y Astronom\'ia, Facultad de Ciencias, Universidad de Valpara\'iso, Gran Bretana N 1111, Playa Ancha, Valpara\'iso, Chile}
\altaffiltext{12}{N\'ucleo de Astronom\'ia de la Facultad de Ingenier\'ia, Universidad Diego Portales, Av. Ej\'ercito 441, Santiago, Chile}
\altaffiltext{13}{Center for Relativistic Astrophysics, School of Physics, Georgia Institute of Technology, Atlanta, GA 30332, USA}
\altaffiltext{14}{Department of Astronomy \& Astrophysics, 525 Davey Lab, The Pennsylvania State University, University Park, PA 16802, USA}
\altaffiltext{15}{INAF -- Osservatorio Astronomico di Bologna, Via Ranzani 1, 40127 Bologna, Italy}
\altaffiltext{16}{Department of Physics, Durham University, South Road, Durham DH1 3LE, UK}
\altaffiltext{17}{School of Physics and Astronomy, University of Southampton, Highfield, Southampton SO17 1BJ, UK}
\altaffiltext{18}{Jet Propulsion Laboratory, California Institute of Technology, Pasadena, CA 91109, USA}
\altaffiltext{19}{Departamento de Astronom\'{\i}a, Universidad de Concepci\'{o}n, Concepci\'{o}n, Chile}
\altaffiltext{20}{SNSF Ambizione Postdoctoral Fellow}
\altaffiltext{21}{NASA Postdoctoral Program Fellow}

\begin{abstract}
We present new \nustar and \Chr observations of NGC 3393, a galaxy reported to host the smallest separation dual AGN resolved in the X-rays.   While past results suggested a 150 pc separation dual AGN, three times deeper \Chr imaging, combined with adaptive optics and radio imaging suggest a single, heavily obscured, radio-bright AGN.  Using VLA and VLBA data, we find an AGN with a two-sided jet rather than a dual AGN and that the hard X-ray, UV, optical, NIR, and radio emission are all from a single point source with a radius $<$0.2\arcsec.  We find that the previously reported dual AGN is most likely a spurious detection resulting from the low number of X-ray counts ($<$160) at 6-7~keV and Gaussian smoothing of the data on scales much smaller than the PSF (0.25$\arcsec$ vs. 0.80$\arcsec$ FWHM).  We show that statistical noise in a single \Chr PSF generates spurious dual peaks of the same separation ($0.55\pm0.07\arcsec$ vs. 0.6\arcsec) and flux ratio ($39\pm9\%$ vs. 32\% counts) as the purported dual AGN.  With {\it NuSTAR}, we measure a Compton-thick source ($N_{\rm H}=2.2\pm0.4\times10^{24}$~\pcmsqsh) with a large torus half-opening angle, $\theta_{\rm tor}=79_{-19}^{+1}\degree$ which we postulate results from feedback from strong radio jets. This AGN shows a 2-10~keV intrinsic to observed flux ratio of $\approx$150 ($L_{2-10 \: \mathrm{keV \: int}}=2.6\pm0.3\times10^{43} \ergps$ vs.~$L_{2-10 \: \mathrm{keV \: observed}}=1.7\pm0.2\times10^{41} \ergpssh$).   Using simulations, we find that even the deepest \Chr observations would severely underestimate the intrinsic luminosity of NGC 3393 above $z>0.2$, but would detect an unobscured AGN of this luminosity out to high redshift ($z\approx5$). 
\end{abstract}



\section{Introduction}
	The detection and measurement of the frequency of dual active galactic nuclei (AGN) is an important test of the merger-driven AGN model.  Inspired by the work of \citet{Sanders:1984:182}, theorists invoke major galaxy mergers as the trigger for both substantial starbursts and quasar phases that sweep galaxies clear of gas, thus starving both star formation and black holes of their fuel \citep[e.g.,][]{DiMatteo:2005:604}.  While secular processes rather than mergers dominate in lower-luminosity AGN \citep[e.g.,][]{Li:2006:457,Schawinski:2011:L31,Treister:2012:L39}, more powerful AGN phases have been seen to occur in major, gas-rich mergers in both nearby \citep{Koss:2010:L125,Koss:2011:57} and high-redshift galaxies \citep{Treister:2010:600}.  Currently, there is significant debate about whether mergers play a significant role in AGN activity particularly at high redshift  \citep[e.g., ][]{Cisternas:2011:57,Kocevski:2012:148,Schawinski:2012:L61,Brandt:2015:1}.  Since AGN are rare, studies of dual AGN, closely separated galaxy nuclei ($<$30 kpc) both hosting AGN in an ongoing merger based on closeness in redshift and disturbed morphologies, have played an important role in studying the frequency of merger-driven AGN activation.  Over the last decade, dozens of these dual AGN on kpc scales have been found serendipitously in interacting galaxies using X-ray imaging, optical spectroscopy, or radio observations \citep[e.g.][]{Komossa:2003:L15,Barth:2008:L119, Comerford:2009:L82, Koss:2011:L42,Fu:2011:L44} or using large-scale surveys in the optical or X-ray \citep{Liu:2011:101,Koss:2012:L22}.  However, despite extensive effort \citep{Iwasawa:2011:106, Burke-Spolaor:2011:2113} and simulations suggesting a peak of  luminous dual AGN activity at small separations   \citep[$<$10 kpc, ][]{VanWassenhove:2012:L7}, few definitive sub-kpc dual AGN have been found. 
	
	As the final phase in a major merger, many models suggest two heavily obscured AGN should form \citep{Hopkins:2009:599}.  NGC 6240, the prototype of a close dual AGN detected in the X-rays over 10 years ago, hosts two Compton-thick AGN separated by 1.5 kpc \citep{Komossa:2003:L15}.   The spectral shape of the 10-100~keV band is especially useful for modeling such heavily obscured, Compton-thick AGN because the 10-20~keV emission will decrease while higher energy emission experiences less absorption when the AGN is in the Compton-thick ($>$$10^{24}$ \pcmsq) to heavily Compton-thick ($5\times10^{24} \pcmsq$) regime.  With the new focusing optics on the {\it Nuclear Spectroscopic Telescope Array} \citep[{\it NuSTAR};][]{Harrison:2013:103}, the 10-79 keV energy range can be studied at sensitivities more than 100$\times$ higher than previous coded aperture mask telescopes such as the {\it Swift} Burst Alert Telescope (BAT) or {\it INTEGRAL}.  
	
	NGC 3393 was reported to be a Compton-thick dual AGN, with the smallest physical separation (150 pc, 0.6\arcsec) ever observed using X-ray imaging \citep{Fabbiano:2011:431}.  However, a recent study using adaptive optics (AO) on Subaru in the $K$-(2.2$\,\micron$) and $L^\prime$ ($3.8\,\micron$) bands found no evidence of a secondary nucleus at 0.6$\arcsec$ despite having higher resolution ($\approx$0.2\arcsec) than \Chr~\citep{Imanishi:2014:106}.  Additionally, an analysis of the nuclear region of NGC 3393 with the VLT Imager and Spectrometer for mid Infrared (VISIR) in four different $N$-band filters ($8-13\,\micron$) found only one compact source at a resolution of 0.3\arcsec \citep{Asmus:2014:1648}.  NGC 3393 was observed by \nustar as part of a survey to observe the most obscured nearby AGN detected with the \swiftbat based on their spectra above 10~keV (Koss et al.~in prep).

	 The host galaxy of NGC 3393 is a face-on spiral galaxy with a Seyfert 2 nucleus \citep{Veron-Cetty:1986:241}.  The nucleus shows polarized broad emission lines \citep{Kay:2002:646}.  Radio images of the inner-kiloparcsec region reveal a core plus an apparent double-sided jet \citep{Cooke:2000:517} with a total extent of $\approx$700 pc.   The radio emission is surrounded by the S-shaped \oiii emission from the narrow line region (NLR) imaged with the Hubble Space Telescope   {\em Hubble Space Telescope\ } \citep[{\em HST},][]{Schmitt:2001:199}. This is consistent with the jets creating denser regions of gas on their leading edges.  A detailed examination of the NLR based on ground-based and space-based imaging and spectroscopy, as well as radio observations, is presented by \citet{Cooke:2000:517}.  The NGC 3393 nucleus is also a source of water maser emission \citep{Kondratko:2006:136,Kondratko:2008:87}.   \bepposax observations in 1997 suggested columns of $N_H=3\times10^{23}$ \pcmsq, but the large Fe K$\alpha$ line equivalent width, high ratio of \oiii to soft X-ray  flux, and the excess above 20~keV suggest a Compton-thick AGN \citep{Salvati:1997:L1}.  The column was later found with \XMM and  \bepposax data to be $N_H=4\times 10^{24}$ \pcmsq \citep{Guainazzi:2005:119} implying that NGC 3393 is Compton-thick, but not so obscured to suppress the nuclear emission fully.  
	 
	In this article, we test for the significance of a dual AGN as well as measure the strength of the AGN obscuration and reflection in NGC 3393.  Since the original NGC 3393 study  where a dual AGN was purportedly detected in the 6-7 keV range \citep{Fabbiano:2011:431}, there are 340 ks of recently obtained \Chr observations which increase the depth of 6-7~keV imaging by over a factor of three.  In addition, there are recently obtained high-resolution milliarcsecond Very Long Baseline Array (VLBA) observations.  Section 2 describes our imaging, spectra, and simulations as well as the analysis technique.  We discuss the host galaxy optical and radio morphology in Section 3.1.   Our results then focus on using multiwavelength data to test whether the source is indeed an X-ray detected dual AGN (Section 3.2).    We then discuss our results from fitting AGN models to NGC 3393 and their implications for the geometry of the torus (Section 3.3).  Finally, a summary and discussion of the difficulty of detecting Compton-thick AGN like NGC 3393 and its importance for higher redshift surveys is discussed (Section 4).  Throughout this work, we adopt $\Omega_m$= 0.27, $\Omega_\Lambda$= 0.73, and $H_0$ = 71 km s$^{-1}$ Mpc$^{-1}$.   We assume a redshift for NGC 3393 of 0.0125 based on 21-cm neutral hydrogen line measurements corresponding to a distance of 53 Mpc \citep{Theureau:1998:333}.  At this redshift, 1$\arcsec$ subtends 250 pc.

\section{Data and Reduction}
Here we present an analysis of new imaging and grating spectroscopy from the X-rays from \nustar (Section 2.1) and \Chr (Section 2.2).  We have also done a reanalysis of past X-ray observations of NGC 3393 (Section 2.3).  A summary of the X-ray observations can be found in Table \ref{tab:xrayobs}. Additionally, we have obtained near-infrared AO imaging (Section 2.4) and optical spectroscopy (Section 2.5).  Finally, we have also analyzed recently obtained VLBA observations and archival VLA radio observations (Section 2.6).  Errors are quoted at the 90\% confidence level unless otherwise specified.

\begin{center}
\begin{table}[tbh]
\scriptsize
\caption{\protect{X-ray Observations of NGC 3393}} 
\label{tab:xrayobs}
\begin{tabular} {cccccc} \hline
Date & Obsid & Telescope & Instrument & Exp.\tablenotemark{1} & Energy\tablenotemark{2} \\
 & &  & & (ks) & (keV) \\
    \hline
1997-01-08 & 50035003 & {\it BeppoSAX} & MECS& 14.6& 0.5-10\\
1997-01-08 & 50035003 & {\it BeppoSAX} & PDS & 14.6& 15-150 \\
2003-07-05 & 140950601 & {\it XMM} & PN, MOS & 15.8& 0.5-10 \\
2004-02-28 & 4868 & {\it Chandra} & ACIS-S & 29.7&  0.5-8 \\
2007-08-14 & 702004010 & {\it Suzaku} & PIN & 55.2&  15-40 \\
2011-03-12 & 12290 & {\it Chandra} & ACIS-S & 67.2&  0.5-8 \\
2012-02-29 & 13967 & {\it Chandra} & ACIS-S HEG & 179.8& 0.5-8 \\
2012-03-06 & 14403 & {\it Chandra} & ACIS-S HEG & 79.0& 0.5-8 \\
2012-04-02 & 14404 & {\it Chandra} & ACIS-S HEG & 57.7& 0.5-8\\
2012-04-08 & 13968 & {\it Chandra} & ACIS-S HEG & 28.5& 0.5-8 \\
2013-01-28 & 61205001 & {\it NuSTAR} & FPMA/B& 15.6 & 3-70 \\
2004-2010 & 70 month & {\it Swift} & BAT & 7333 & 14-195\\
        \hline 
   \end{tabular}
\footnotetext[1]{Effective exposure time (ks) after data cleaning and correction for vignetting. }
\footnotetext[2]{Energy range studied in analysis. }
\end{table}
\end{center}

\subsection{\nustar}
The raw data were reduced using the {\tt NuSTARDAS} software package (version 1.3.1) jointly developed by the ASI Science Data Center (ASDC) and the California Institute of Technology. {\tt NuSTARDAS} is distributed with the HEAsoft package by the NASA High Energy Astrophysics Archive Research Center (HEASARC).  We extracted the \nustar source and background spectra using the {\tt nuproducts} task with the appropriate response and ancillary files.  Spectra were extracted from circular regions 40$\arcsec$ in radius, centered on the peak of the centroid of the point-source. The background spectra were extracted from three regions on the same detector as the source.    The final exposure time after screening was 15.7 ks,  with 821 and 758 background subtracted counts, respectively,  for the FPMA and FPMB modules.   The \nustar spectra are binned to a minimum of 20 photons per bin using HEAsoft task {\tt grppha}.

\subsection{\Chr}
\begin{figure} 
\includegraphics[width=8cm]{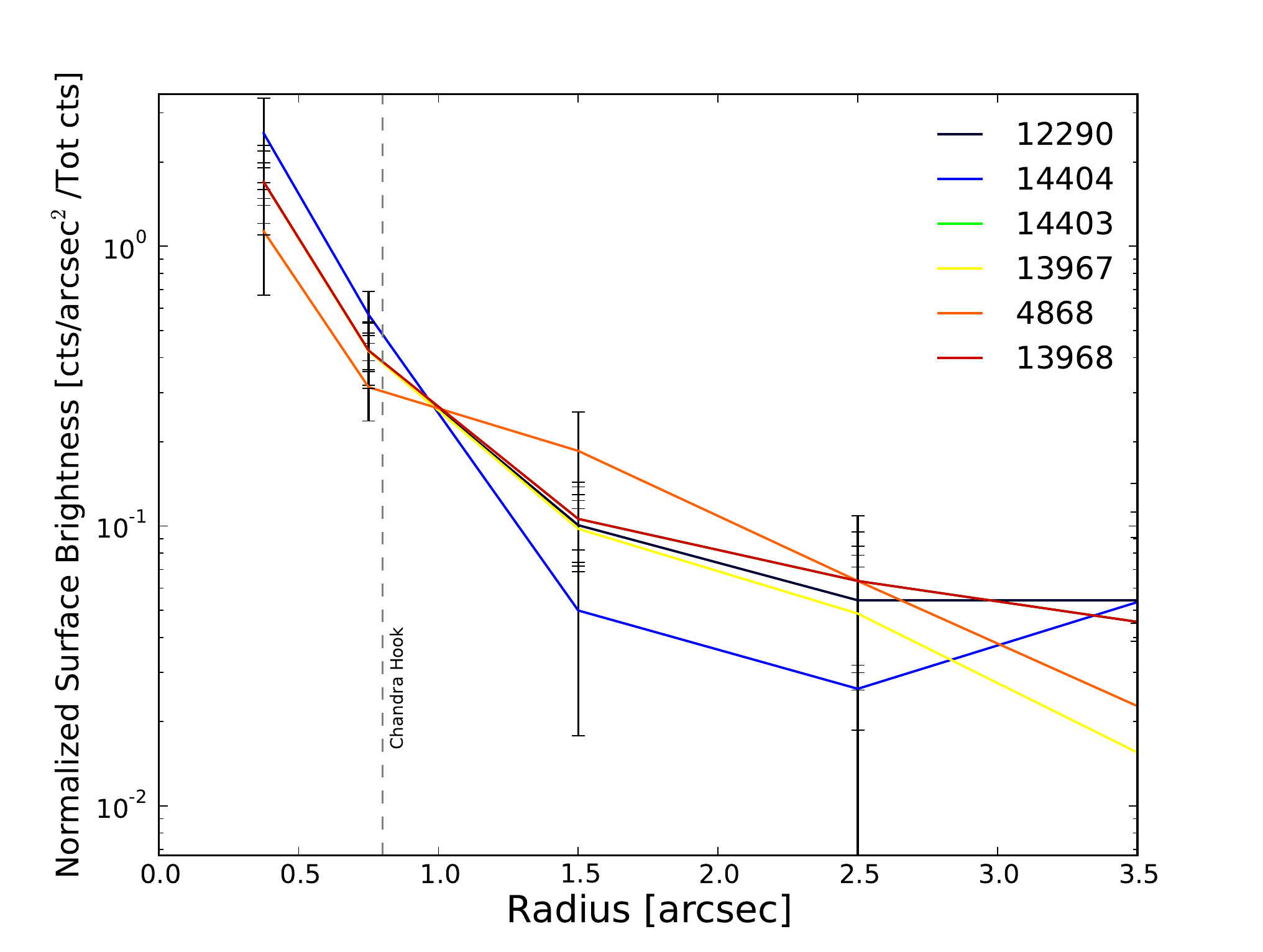}
\caption{Radial profiles of the ACIS images (12290 and 4868) and zeroth order grating data (13967, 13968, 14403, 14404).  A dashed line indicates the position of the \Chr hook asymmetry.  We see no statistically significant evidence that any of the images are different from each other in terms of their PSF size.}
\label{grating_comp}
\end{figure}

Archival \Chr Advanced CCD Imaging Spectrometer (ACIS) X-ray imaging taken on 2004 February 28  (ObsID 4868 for 30 ks total exposure) and 2011 March 12 (ObsID 12290 for 70 ks total exposure) is used in the analysis.  The  effective area of the \Chr High Energy Transmission Grating \citep[HETG,][]{Canizares:2000:L41} zeroth-order image is about 60\% of the effective area of ACIS-S at 6.4~keV, so we also make use of 340~ks of recent HETG observations taken in 2012 (PI: Evans; ObsID 13967, 170 ks; ObsID 13968, 30 ks; ObsID 14403, 79 ks; ObsID 14404, 61 ks), providing an equivalent of 210 ks ACIS-S imaging in the Fe K$\alpha$ band ($\approx$6-7 keV).  Therefore, the combined depth of our imaging data reaches 310 ks, which improves the exposure by a factor of 3.1 in the 6-7~keV range   where a dual AGN was purportedly detected in a past study of NGC 3393 \citep[100 ks depth;][]{Fabbiano:2011:431}.  The total number of counts at 6-7~keV in a 2\arcsec radius aperture increases from 145 to 449, a factor of 3.1, consistent with the increase in exposure time.  We find no evidence that the imaging point spread function (PSF) is significantly worse in any of the ACIS images or zeroth order grating data at 6-7 keV (Figure \ref{grating_comp}).

The data were analyzed using {\tt CIAO} (Version 4.5) with the latest CALDB 4.5.8 provided by the \Chr X-ray Center (CXC). The subpixel positioning was applied using Energy Dependent Subpixel Event Repositioning \citep[EDSER,][]{Tsunemi:2001:496,Li:2003:586,Li:2004:1204} with {\tt CIAO} tool {\tt acis\_process\_events}.  The events were screened for high-background periods with the {\tt CIAO} tool {\tt deflare} with a 2.5$\sigma$ cutoff. Spectral responses were generated using the {\tt CIAO} tool {\tt specextract}.  

In order to correct for offsets in the relative astrometry of each \Chr observation, we use the CIAO tools {\tt fluximage} to create exposure-corrected images and exposure maps, {\tt mkpsfmap} to compute the point spread function (PSF) size across the image, and {\tt wavdetect} for source detection.  We include only the highest significance point sources when computing astrometry (SN$>$10, PSF size$<$1.5$\arcsec$), and detect seven sources in both images, excluding the nucleus.  We find an offset of 0.41$\arcsec\pm$0.08$\arcsec$ in R.A. and 0.11$\arcsec\pm$0.05$\arcsec$ in decl. between ObsID 4868 and 12290.   We can also compute the relative astrometric offset using the 4-8~keV centroid of NGC 3393 and find similar values with the same offset direction (0.50$\arcsec$ in R.A. and 0.07$\arcsec$ in decl. at 4-8~keV and 0.45$\arcsec$ in R.A. and 0.12$\arcsec$ in decl. at 6-7~keV).   This is consistent with the 90\% uncertainty (0.63$\arcsec$) in ACIS-S astrometry of on-axis sources\footnote{See {\tt http://cxc.harvard.edu/cal/ASPECT/celmon/} for more details on astrometry.}.  All seven sources show the same offset direction in both R.A. and decl. We then use the CIAO tool {\tt reproject\_aspect} to align images to the same world coordinate system (WCS).  In images taken with the HETG offset sources are not available for crossmatching, so we combine the images based on the centroid of NGC 3393.

NGC 3393 has evidence of extended emission, so we extract a nuclear region (2$\arcsec$ radius) enclosing the PSF and a 40$\arcsec$ radius matched to the \nustar observation. There is no evidence of pileup in any of the \Chr imaging or grating observations because of the low count rate of the nuclear emission ($\approx0.015$~\cps in ACIS-S imaging).  The nuclear region is corrected for the small amount of PSF flux ($<10\%$) beyond the extraction radius.

We also use the HETG Medium Energy Grating (MEG) and High Energy Grating (HEG) data from recently obtained archival data with a total exposure time of 340 ks.  The gratings operate simultaneously, with the MEG/HEG dispersing a fraction of the incident photons from the two outer/inner High Resolution Mirror Assembly (HRMA) shells.    The 1st-order HETG spectral products were extracted using the {\tt tgcat} pipeline for grating spectra in {\tt ISIS}.  We use a full-width of 4$\arcsec$ in the cross-dispersion direction of NGC 3393 and extract it as a point source.  This corresponds to a maximum distance from the central source of $\approx$500 pc.  We find that the count rates in the MEG, HEG, and Fe K region for observations 13967, 13968, and 14403 are all within 1$\sigma$ of each other.  Observation 14404 has a larger difference, with a count ratio of 1.8$\sigma$ lower than the average of the others, which we do not consider significant.  We therefore coadd all the spectra.  The grating spectra then have a total of 1259 counts in the MEG and 736 counts in the HEG and are fit with models simultaneously to improve the fit.     We note that NGC 3393 is extended in the soft X-rays ($<$ 3 keV) causing a degradation of the apparent spectral resolution.  

\subsection{Other X-ray Telescopes}
We also include data from other X-ray observatories because of their overlap with \nustar and to provide constraints on variability between the energy range of 0.5-200 keV.  A summary of these additional X-ray observations can be found in Table \ref{tab:xrayobs}.

NGC 3393 was observed by \bepposax on 1997 January 8  with the Low Energy Concentrator Spectrometer (LECS), the three Medium Energy Concentrator Spectrometers (MECS), and the Phoswich Detector System  (PDS). There was no detection in the LECS, so we use only the MECS and PDS here.  The MECS contains three identical gas scintillation proportional counters, with angular resolution of $\approx$0.7$\arcmin$ full width half max (FWHM) and $\approx$2.5$\arcmin$ half power diameter (HPD). The MECS event files were screened adopting standard pipeline selection parameters.  Spectra were extracted from 4$\arcmin$ radii apertures and the spectra from the three units were combined after renormalizing to the MECS1 energy-PI relation  covering the energy range 0.5-10 keV.  The PDS has no imaging capability to speak of, but has sensitivity to 15--150~keV with some overlap with \nustarsh.  The PDS data were calibrated and cleaned using the {\tt SAXDAS} software, adopting the fixed Rise Time threshold method for background rejection.

NGC 3393 was observed by \suzaku with the Hard X-ray Detector (HXD) PIN on 2007 August 14 with the nominal HXD pointing.  We reprocessed the unfiltered events file using the standard \suzaku pipeline FTOOL {\tt aepipeline}.  The FTOOL {\tt hxdpinxbpi} was used to extract the source spectrum and to create a total background spectrum from the tuned non-X-ray background for the NGC 3393 observation provided by the \suzaku team.  The filtered events file, corrected for dead time, has a total exposure of 45 ks covering the energy range 14-50 keV.  The \suzaku source spectrum was binned such that, in each bin, the source is three times the background.  

 NGC 3393 was observed by \XMM on 2003 July 5. We processed \XMM data using SAS (v13.5.0) and the {\tt xmmextractor} metatask for end to end processing of all the \XMM raw data to science level data with filtering.  This left 13.1 ks usable exposure.  The background events are extracted from a annulus of inner radius 70$\arcsec$ and outer radius 140$\arcsec$ centered on the source position.  We processed the EPIC MOS and PN data specifically using standard values for a point source.   We use a 40$\arcsec$ radius matched to the \nustar observation for extraction and use the 0.5-10 keV range in EPIC MOS and PN data.  There is no evidence of pileup in the observation.


We finally use the \swiftbat stacked spectra and light curve from the publicly available 70 month catalog which spans 2004--2010 covering 15-195 keV.  The light curve data was binned into 3 month intervals.  Details of the BAT catalog reduction procedure can be found in \citet{Baumgartner:2013:19}.

When comparing observations between X-ray telescopes to assess variability it is important to consider the much wider PSFs of other instruments because of contamination by other sources as well as cross-calibration before comparing them with $NuSTAR$.   This is especially important for \bepposax data because of its lower angular resolution.   For instance, the \bepposax MECS data has a beam size of 4$\arcmin$.  Based on \XMM image, we find that other AGN separated between 0.9$\arcmin$ and 4$\arcmin$ from NGC 3393 have $\approx$45\% of the 4.5-12~keV counts.  The \bepposax PDS FWHM is a large 1.3$\degree$.  Based on the higher resolution \swiftbat maps, there is another source $\approx$0.88$\degree$ NE detected at SNR=3.75 at R.A.=162.656$^{\circ}$ and decl=-24.306$^{\circ}$, which is near (2.8$\arcmin$ offset) the source 1RXS J105027.3-241644.  The source has a count rate roughly 29\% of counts of NGC 3393 based on the \swiftbat 14-195~keV maps (0.000327~\cps~vs. 0.000796~\cps).

We assume a cross-calibration of 0.93 between \suzaku PIN and \nustar and 1.25 between \XMM and \nustar based on recent tests of other calibration sources.  These values are based on the   current best estimates for \nustar (Madsen et al., submitted).  For \bepposax MECS we assume a factor of 0.55 based on the contamination and a further cross-calibration of 0.8 between MECS and PDS following the instrument guidelines.  We have assumed a value of 1 for the cross-calibration of \nustar compared to {\tt Swift} BAT  because \swiftbat takes several years to reach the same significance as \nustarsh and has not yet been well calibrated (K. Madsen, 2015, Private Communication).

\subsection{High Resolution Optical and Near Infrared Imaging}

We imaged NGC 3393 on 2013 November 10 in the near infrared (NIR) using AO in laser guide star mode with the Near Infrared Camera 2 (NIRC2) instrument on the Keck-2 telescope. We used the wide camera with 40 mas pixel$^{-1}$  and a 40$\arcsec$ field of view (FOV).   We used a 3 point dither pattern for 18 minutes total in the $K_p$ filter.  Images were combined using a Strehl weighting with Strehl ratios between 5-7\% for a nearby star.  The final PSF (FWHM) is 0.15\arcsec, corresponding to physical scales of $\approx$40 pc, as measured from nearby stellar sources in the final co-added image.  The absolute astrometry of the NIR image was recalibrated using three bright stars from a PS1 r-band image.

	 We analyzed \hst imaging from GO program 12185 using Wide Field Camera 3 (WFC3) for filters F336W, F438W, and F814W.  We also processed imaging from \hst Advanced Camera for Surveys (ACS) F330W (Program 9379) and narrowband WFC3 FQ508N \oiii  (Program 12365).  We apply AstroDrizzle to flat-field calibrated images to create mask files for bad pixels and cosmic rays.  Additionally, when the observations have multiple images, we drizzle-combine the input images using the mask files, while applying geometric distortion corrections, to create a distortion-free combined image.    The absolute astrometry of each image was finally recalibrated using three bright stars from a PS1 r-band image.

\subsection{Optical Spectroscopy}
We also observed NGC 3393 using the University of Hawaii 2.2m telescope and the SuperNova Integral Field Spectrograph (SNIFS) on 2013 April 27 for a total duration of 3600 seconds.  SNIFS is an optical integral field spectrograph with blue (3000--5200 $\mathrm{\AA}$) and red (5200-9500 $\mathrm{\AA}$) channels and a resolution of 360 $\kmpssh$.  The SNIFS reduction pipeline {\tt SNURP} was used for wavelength calibration, spectro-spatial flat-fielding, cosmic ray removal, and flux calibration \citep{Bacon:2001:23,Aldering:2006:510}.  A sky image was taken after each source image and subtracted from each Integral Field Unit (IFU) observation.  Flux corrections were applied based on the standard star Feige 34.  We fit the spectra using an extensible spectroscopic analysis toolkit for astronomy, {\tt PYSPECKIT}, which uses a Levenberg-Marquardt algorithm for fitting.   

\subsection{Radio Observations}

\begin{figure*}
\plotone{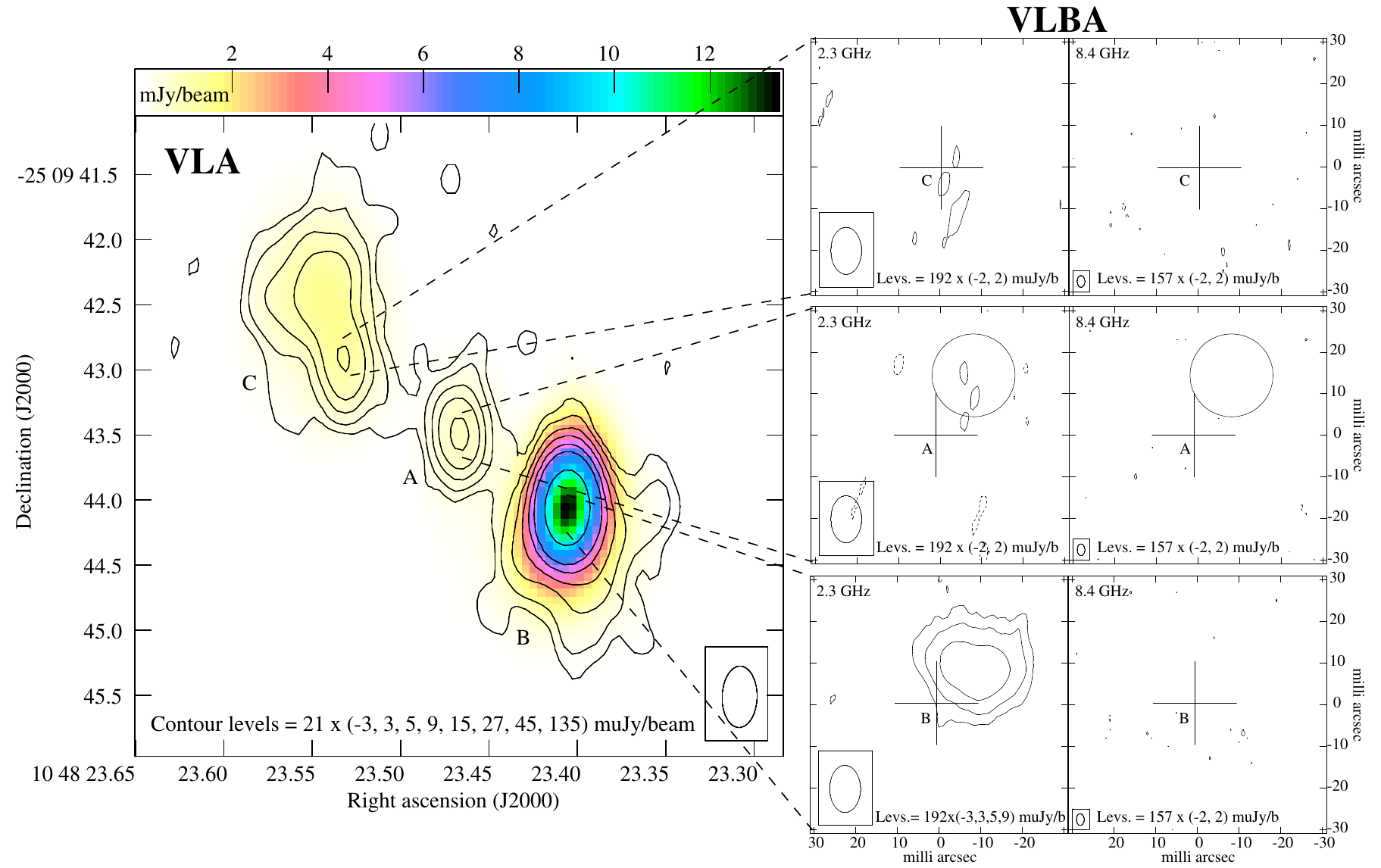}
\caption{NGC 3393 VLA 8.4~GHz contours overlaid on color-scale 4.9~GHz map (left).  Width of the radio image is 4.8$\arcsec$. VLBA 
contour images from each radio component A, B and C at 2.3 and 8.4~GHz (right). The plus
signs in the VLBA images represent the A, B and C peak positions as obtained from the 
VLA 8.4~GHz image, with an accuracy better than 10 mas. The circle in the VLBA images of
component A, represents the dynamical location of the BH obtained from water maser 
emission \citep{Kondratko:2008:87}.
}\label{fig:ngc3393radio}
\end{figure*}

NGC 3393 was observed with the VLA on 1992 November 29 (project code AB618) at 1.4 GHz 
($3.0\times1.5^{\prime\prime}$ beam size, PA$=-14.1^{\circ}$), 4.9~GHz ($0.8^{\prime\prime}\times0.4^{\prime\prime}$ beam size, 
PA$=-7.7^{\circ}$), and 8.4~GHz ($0.5^{\prime\prime}\times0.3^{\prime\prime}$, PA$=-1.5^{\circ}$).  In Figure \ref{fig:ngc3393radio}, we show 8.4 
GHz contours overlaid on the color-scale 4.9~GHz VLA image (left panel).  Three components, referred to as A (nucleus), B (NE jet lobe), and C (SE jet lobe) are clearly detected in agreement with past studies \citep[See Section 1,][]{Cooke:2000:517}.  The positions of the 
different components measured from the VLA 8.4~GHz map (the one with the highest resolution) 
are shown in Table \ref{tab:radio_posn}.  The larger beam in the 1.4 GHz VLA map prevents us from resolving components A, B and C separately, and 
thus we no longer consider it in our analysis.  

We also reduced and combined two days of archival VLBA observations (2011 December 7-8; project 
code BS214) at 2.3 GHz ($11.3\times7.4$\,mas, PA$=0.4^{\circ}$) and 8.4~GHz ($2.8\times1.8$\,mas, 
PA$=1.8^{\circ}$). Whilst A, B, and C are clearly detected in 
the VLA images (Figure 1), only component B is detected in the VLBA image at 2.3 GHz, and appears as an extended 
source ($S_{=2.3 \mathrm{GHz}} = 7.54 \pm 0.42$ mJy). This further suggests B is a jet hotspot 
component. Deeper images are needed to detect components A and C. Alternatively,
observations at higher frequencies can also help in the detection of component A, if this represents
a case similar to 4C39.25, whose core has a spectral turnover at high frequencies, between 43 and 86~GHz 
\citep{Alberdi:1997:513}.

\begin{center}
\begin{table}[tbh]
\centering
\caption{\protect{NGC 3393 radio components from VLA 8.4~GHz observations}} \label{tab:radio_posn}
\begin{tabular} {ccccc} \hline
\multicolumn{1}{c}{Component} & \multicolumn{2}{c}{Peak position} & \multicolumn{1}{c}{$D_{\mathrm{Jet-core}}$} \\
\multicolumn{1}{c}{} &  \multicolumn{1}{c}{RA(J2000)} & \multicolumn{1}{c}{DEC(J2000)} & \multicolumn{1}{c}{(arcsec)}\\
    \hline
A  & 10 48 23.467 & -25 09 43.49 & -    \\
B  & 10 48 23.406 & -25 09 44.06 & 1.00 \\
C  & 10 48 23.532 & -25 09 42.91 & 1.06 \\
    \hline 
   \end{tabular}
\begin{center}
Position uncertainties in R.A.~and decl. are better than $\pm 10$\,mas.
\end{center}
\end{table}
\end{center}

\begin{center}
\begin{table}[tbh]
\centering
\caption{\protect{Flux densities and luminosities from 
VLA observations at matched resolutions.}} \label{tab:radio_measurements}
\begin{tabular} {cccc} \hline
\multicolumn{1}{c}{$\nu$} & \multicolumn{1}{c}{Component} & \multicolumn{1}{c}{$S_{\nu}$} & \multicolumn{1}{c}{$L_{\nu}$} \\
\multicolumn{1}{c}{(GHz)} & \multicolumn{1}{c}{} & \multicolumn{1}{c}{(mJy)} & 
  \multicolumn{1}{c}{($10^{27}$ erg s$^{-1}$ Hz$^{-1}$)} \\
    \hline
8.4 & A &  0.70 $\pm$ 0.04 &  2.36 $\pm$ 0.14 \\
    & B &  9.22 $\pm$ 0.46 & 30.98 $\pm$ 1.55 \\
    & C &  1.92 $\pm$ 0.10 &  6.44 $\pm$ 0.33 \\
\hline
4.9 & A &  0.83 $\pm$ 0.05 &  2.78 $\pm$ 0.17 \\
    & B & 15.16 $\pm$ 0.76 & 50.95 $\pm$ 2.55 \\
    & C &  2.89 $\pm$ 0.15 &  9.71 $\pm$ 0.49 \\
    \hline 
   \end{tabular}
\begin{center}
Uncertainties in flux density include a conservative 5\% error in the point source calibration and the rms noise in 
the map, added in quadrature.
\end{center}
\end{table}
\end{center}

\begin{figure}
\plotone{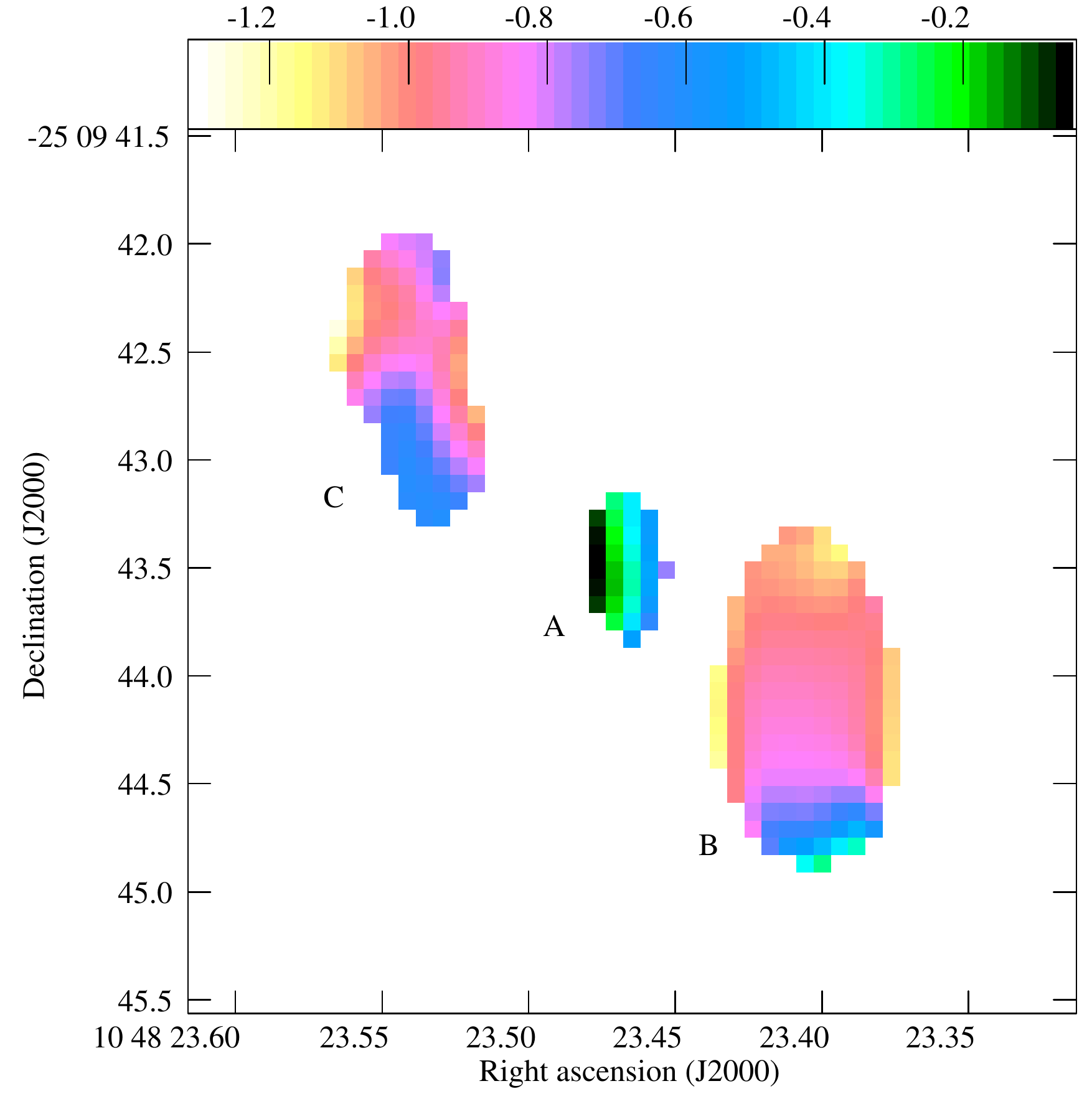}
\caption{Pixel by pixel spectral index distribution between 4.9 and 8.4~GHz emission detected with the VLA.  The structure recovered with the VLA resembles an AGN with a two-sided jet rather than a dual AGN  \citep[both here and in][]{Cooke:2000:517}.  Component A (nucleus), with a flat spectrum, is the core, and the steep spectrum components are the lobes.  Their flux difference explained by Doppler boosting, would be approaching and receding components, respectively.}
\label{fig:radiospix}
\end{figure}

In order to obtain useful spectral index information from the VLA images, we created 
4.9 and 8.4~GHz maps with matched baseline ranges in wavelength, and the same convolving beam
($0.65\arcsec \times0.35\arcsec$, PA$=0^{\circ}$).  We then obtained a spectral index distribution map considering only emission above 15$\sigma$ (see Table \ref{tab:radio_measurements}), to avoid confusion between the different components (Figure~\ref{fig:radiospix}). 

\section{Results}
\subsection{Single or Dual AGN?}
 Before studying the X-ray spectra, we assess the significance of the previously claimed dual AGN.  Our results first focus on the high resolution radio, optical, and NIR morphology of NGC 3393 to test whether their is evidence of the purported dual AGN or a recent merger.  We then test for astrometric offsets in the multi-wavelength data.  Finally, we use deconvolution, modeling, and simulations of \chandra X-ray data to test the significance of the purported dual AGN. 

\subsubsection{Radio Morphology}
The structure recovered with the VLA is indicative of that from an AGN with a two-sided jet rather than a dual AGN.  Component A (nucleus) with a flat spectrum is the core, whereas B and C have steep spectra with a flux difference easily explained by Doppler boosting. In this interpretation, components B and C are the approaching and receding components, respectively (see Figure \ref{fig:radiospix}). Using the flux  density enclosed by the 15$\sigma$ contour in each component, 
we obtain spectral indices $\alpha$=$-0.30\pm0.15$, $-0.90\pm0.13$ and $-0.74\pm0.13$ for components A,  B and C, respectively ($S_{\nu} \propto \nu^{\alpha}$).

\subsubsection{High-Resolution Optical and NIR Morphology}
\begin{figure} 
\plotone{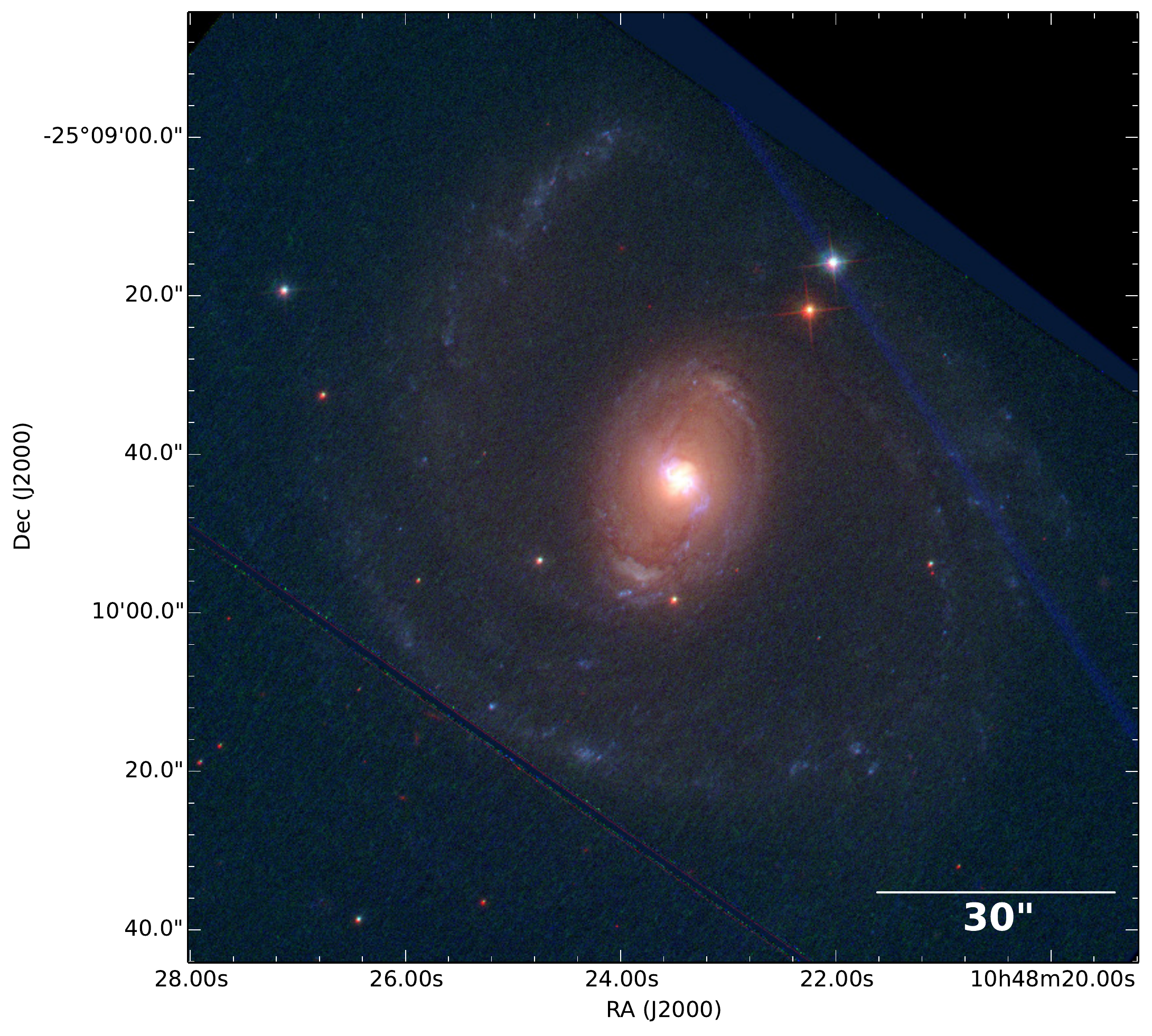}
\caption{Tricolor \hst images (F336W, F438W, F814W) of NGC 3393.  The image is 2$\arcmin$ on a side.  NGC 3393 is a spiral galaxy that shows faint tightly wound spiral arms which are much stronger in the $H\alpha$ image \citep{Cooke:2000:517} that connect to a bar at PA $\approx$160$\degree$.  There is no evidence of any asymmetry in the spiral arms as would be expected from a recent merger.  }
\label{hst_tri}
\end{figure}

	 We use the high-resolution \hst imaging to search for signs of a recent merger.  A \hst tricolor (F336W, F438W, F814W) image can be found in Figure \ref{hst_tri}. NGC 3393 is an early-type barred spiral that is nearly face on \citep[SBars,][]{deVaucouleurs:1995}.  NGC 3393 shows faint, tightly wound, blue spiral arms, which are much stronger in the H$\alpha$ image \citep{Cooke:2000:517}. There is no evidence of any asymmetry in the spiral arms or any sign of tidal tails indicative of a recent merger.

\begin{figure*} 
\includegraphics[width=8.5cm]{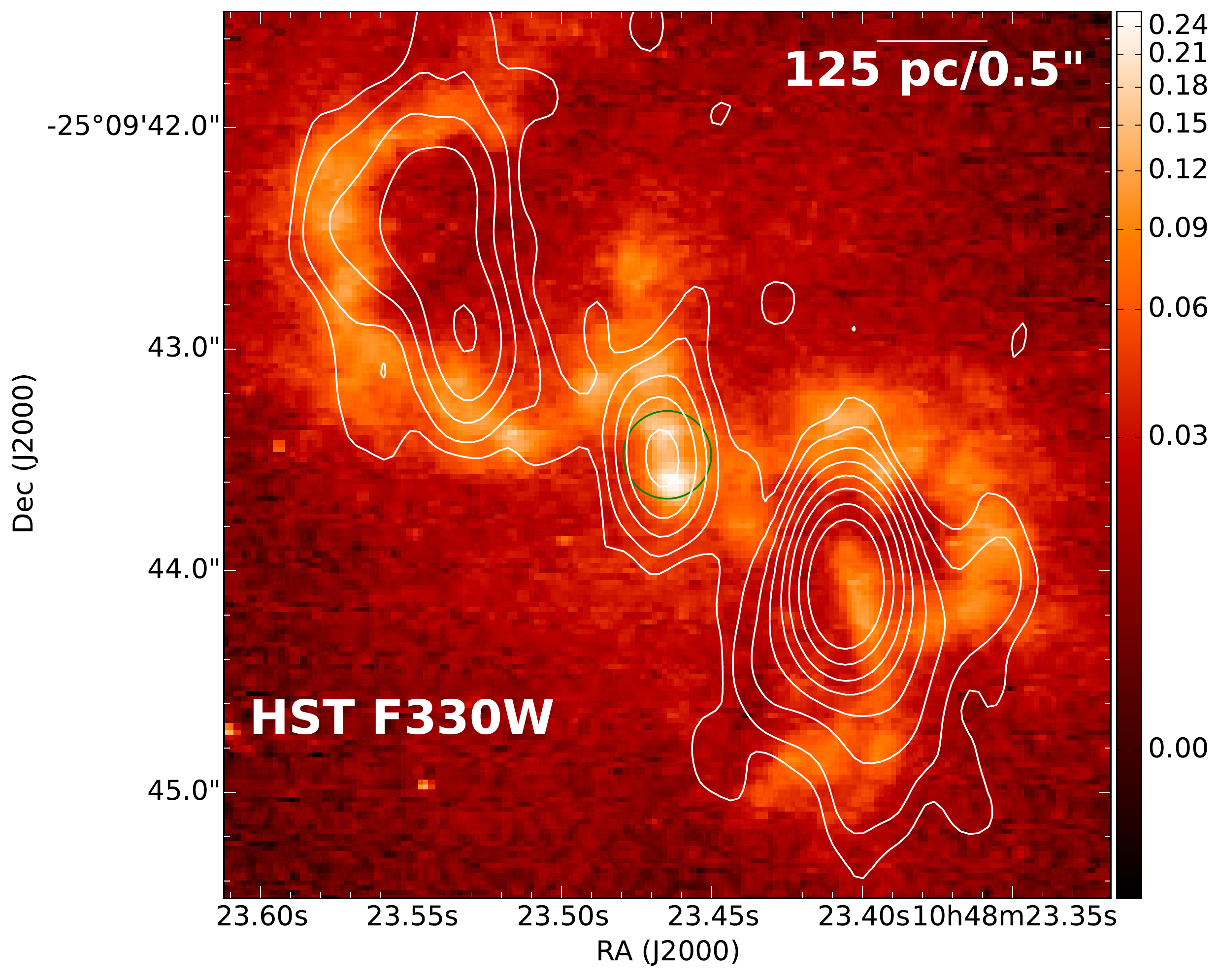}
\includegraphics[width=8.5cm]{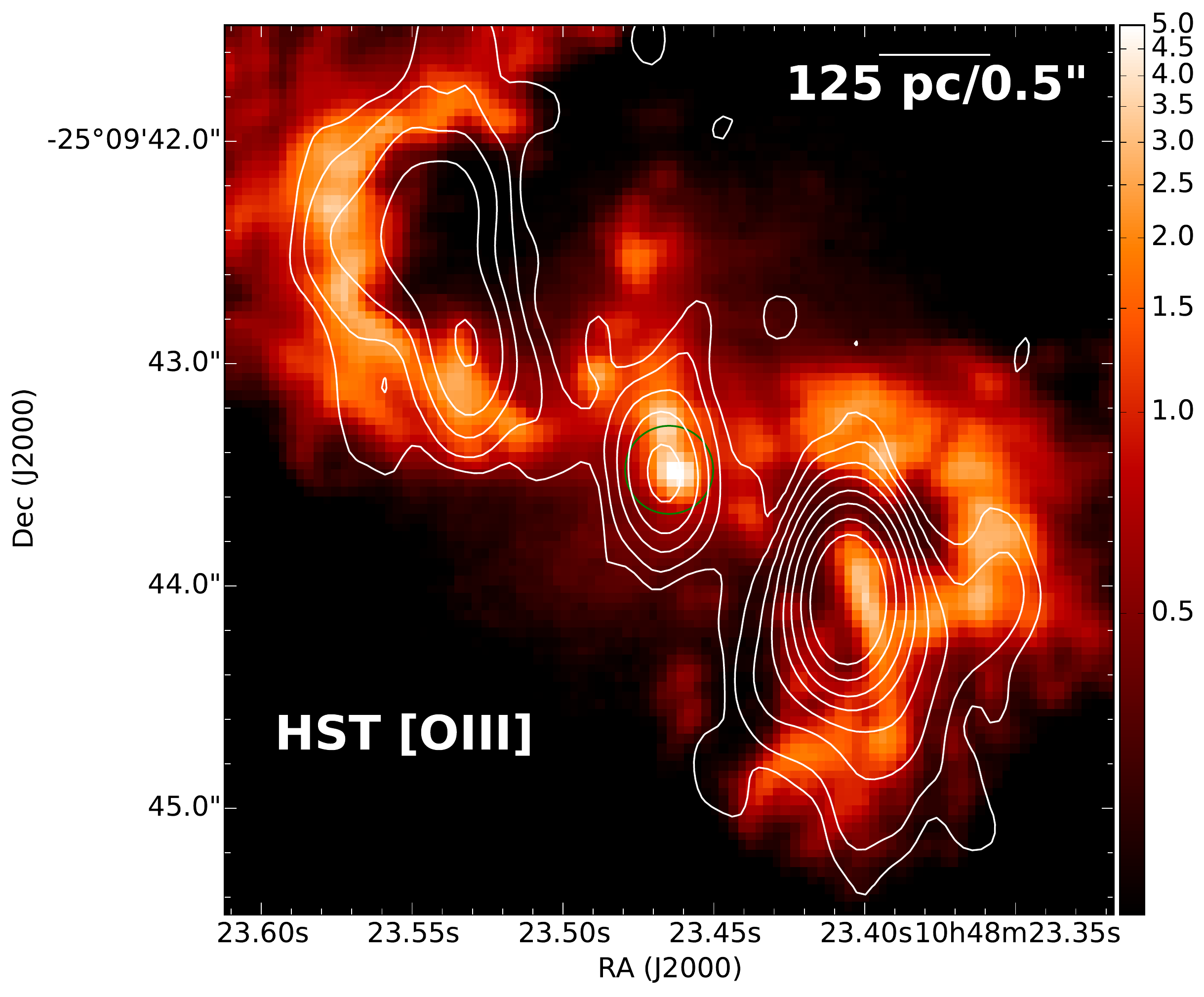}
\includegraphics[width=8.5cm]{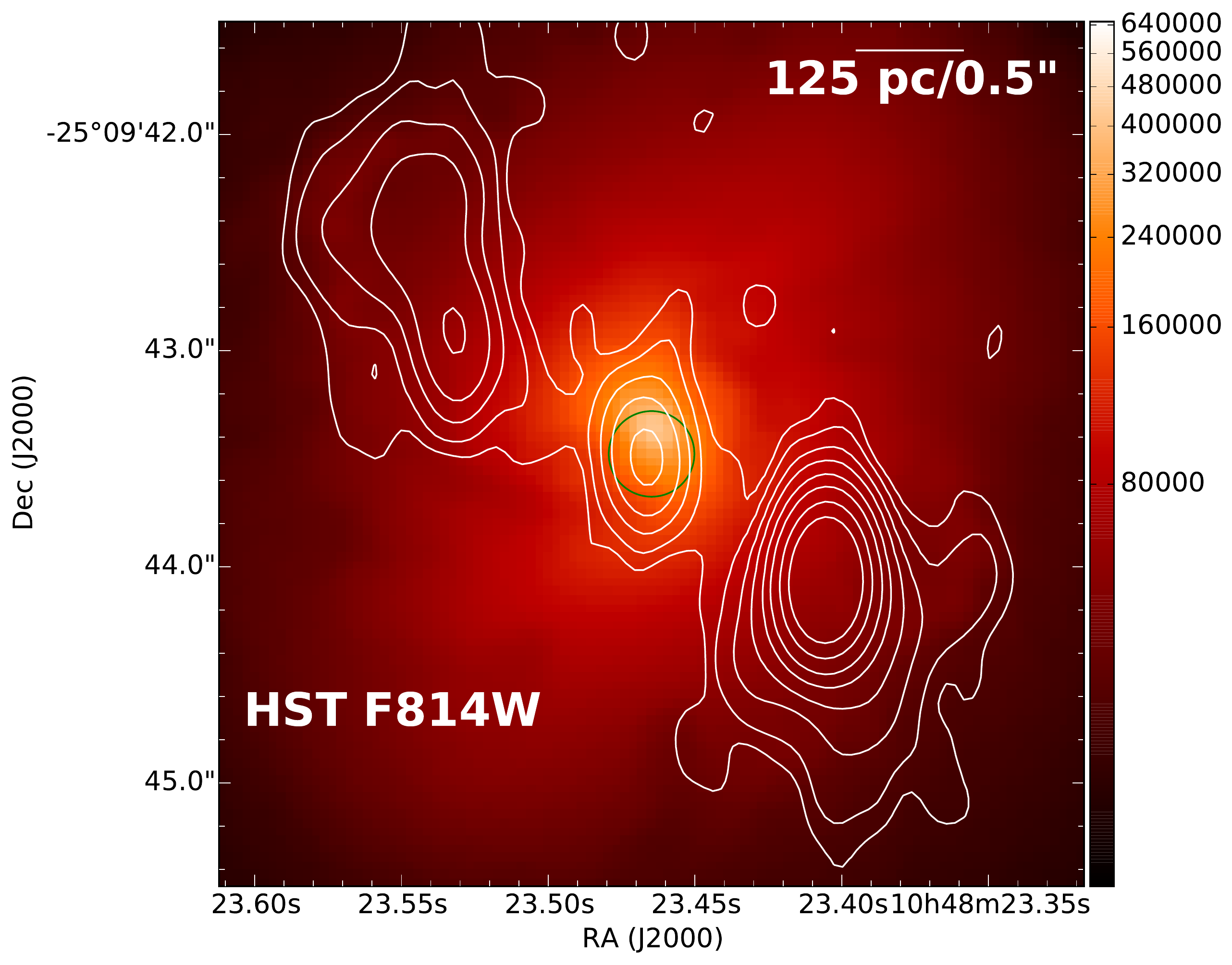}
\caption{\hst image of the nuclear region of NGC 3393 compared to VLA 8.4~GHz radio contours for the 330W, \oiiish, and 814W filters.  A green circle of 0.2$\arcsec$ radius represents the dynamical location of the BH obtained from water maser emission \citep{Kondratko:2008:87}.  In the nucleus an S-shaped source can be seen in the bluer images coincident with the radio emission.  The radio lobes created denser regions of gas on their leading edges similar to a bow-shock \citep{Cooke:2000:517}.  In the F330W UV and \oiii there is evidence of  small biconical region of 0.2$\arcsec$ that is not detected in the redder bands.  There are some small absolute offsets with the radio data, however these offsets are consistent within the 0.2$\arcsec$ astrometric uncertainty.}
\label{hst_overlay}
\end{figure*}

\begin{figure} 
\includegraphics[width=8.5cm]{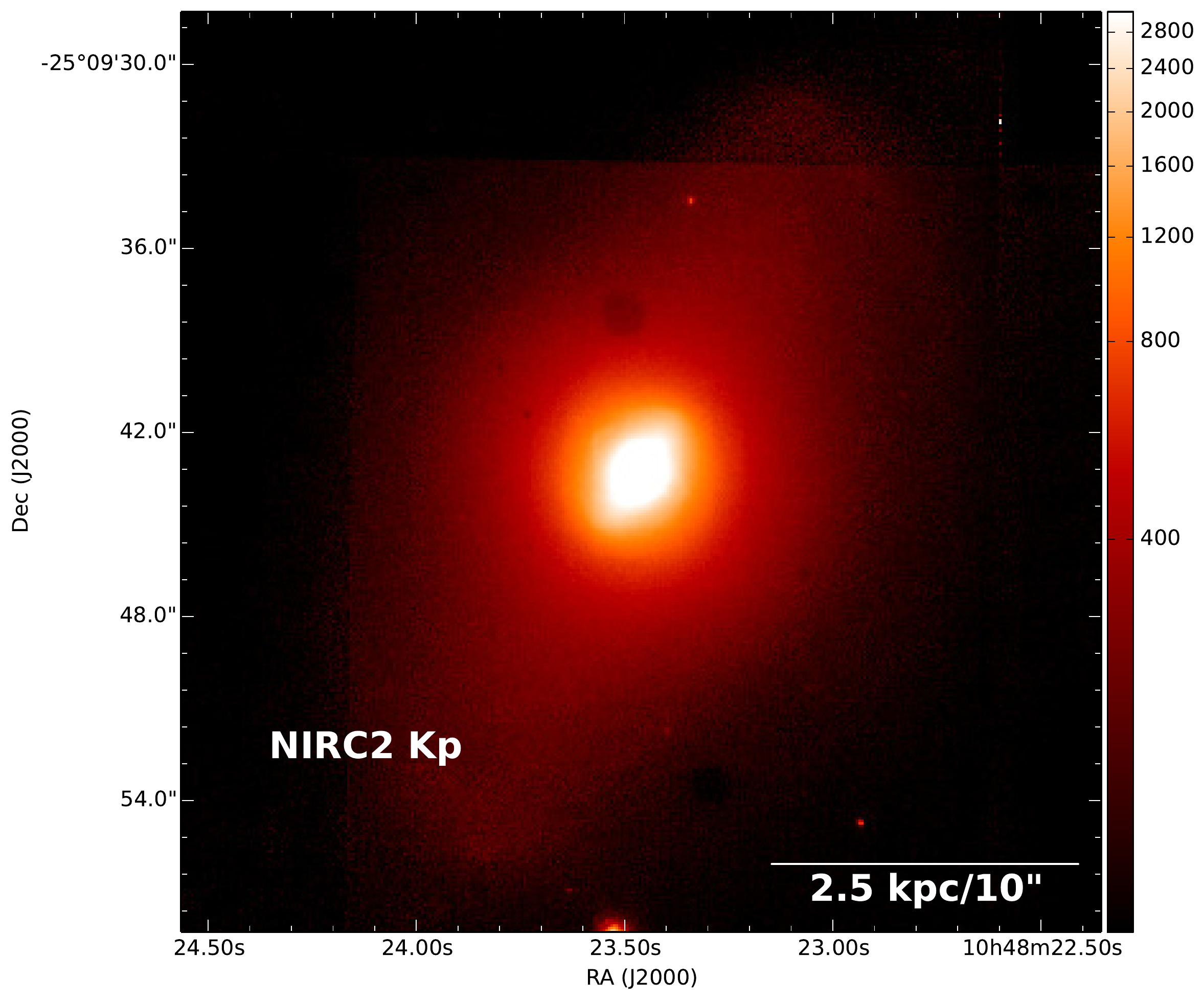}
\includegraphics[width=8.5cm]{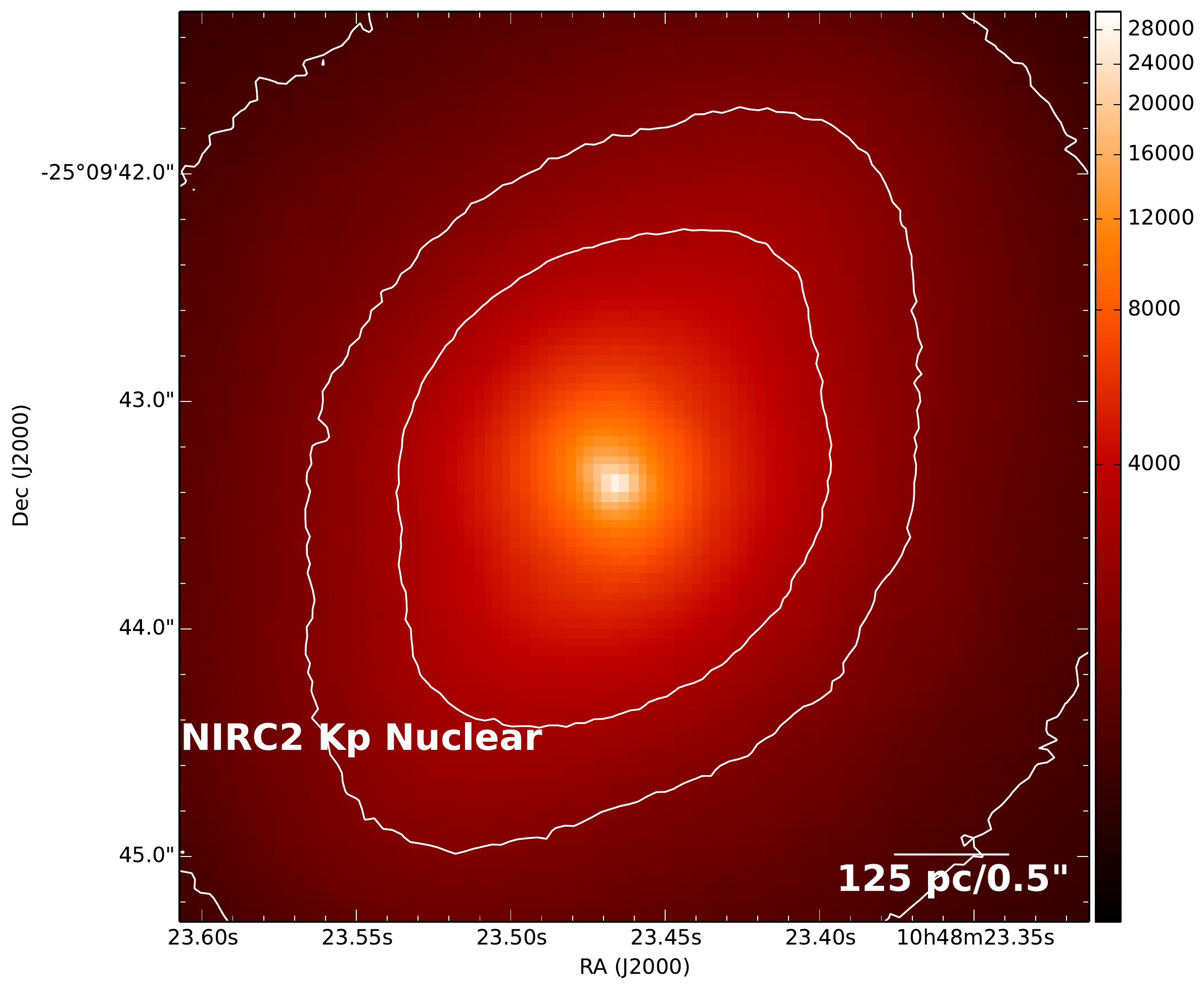}
\caption{{\em Top:} Image of NGC 3393 in the $K^\prime$ band.  The PSF FWHM is 0.10$\arcsec$, corresponding to physical scales of 25 pc.  The images show a bar at PA 160$\degree$.  {\em Bottom:}  Zoom in around NGC 3393.  The diameter of the smallest white contour is $\approx$100 pc.  We do not see any evidence of an obscured secondary nucleus and the contours show a smooth potential consistent with a single nucleus.  }
\label{nirc2}
\end{figure}

	We then compare the \hst images of the nuclear region of NGC 3393 to radio contours for the F330W, [O\,{\sc iii}], and F814W filters (Figure~\ref{hst_overlay}).  In the nucleus, an S-shaped source can be seen in the bluer images coincident with the radio emission.  In the UV and \oiii there is evidence of a small biconical region on the nucleus  that is extended by 0.2\arcsec. The identical biconical structure in the NUV and \oiii suggests that the near-UV emission is produced in the same region as the ionized gas, and thus the nebular continuum and the [NeV] emission must be the main contributors as noted by past studies \citep{MunozMarin:2009:842}.  Additionally, the biconical structure in the \hst UV and \oiii is likely due to star formation or photoionized emission because of their lack of detection in the NIR imaging. Finally, in the NIRC2 images (Figure~\ref{nirc2}), we find a bar at PA$\approx$160$\degree$, with a smooth distribution and no evidence of a secondary nucleus.  
	
\subsubsection{Astrometry}
\begin{figure} 
\plotone{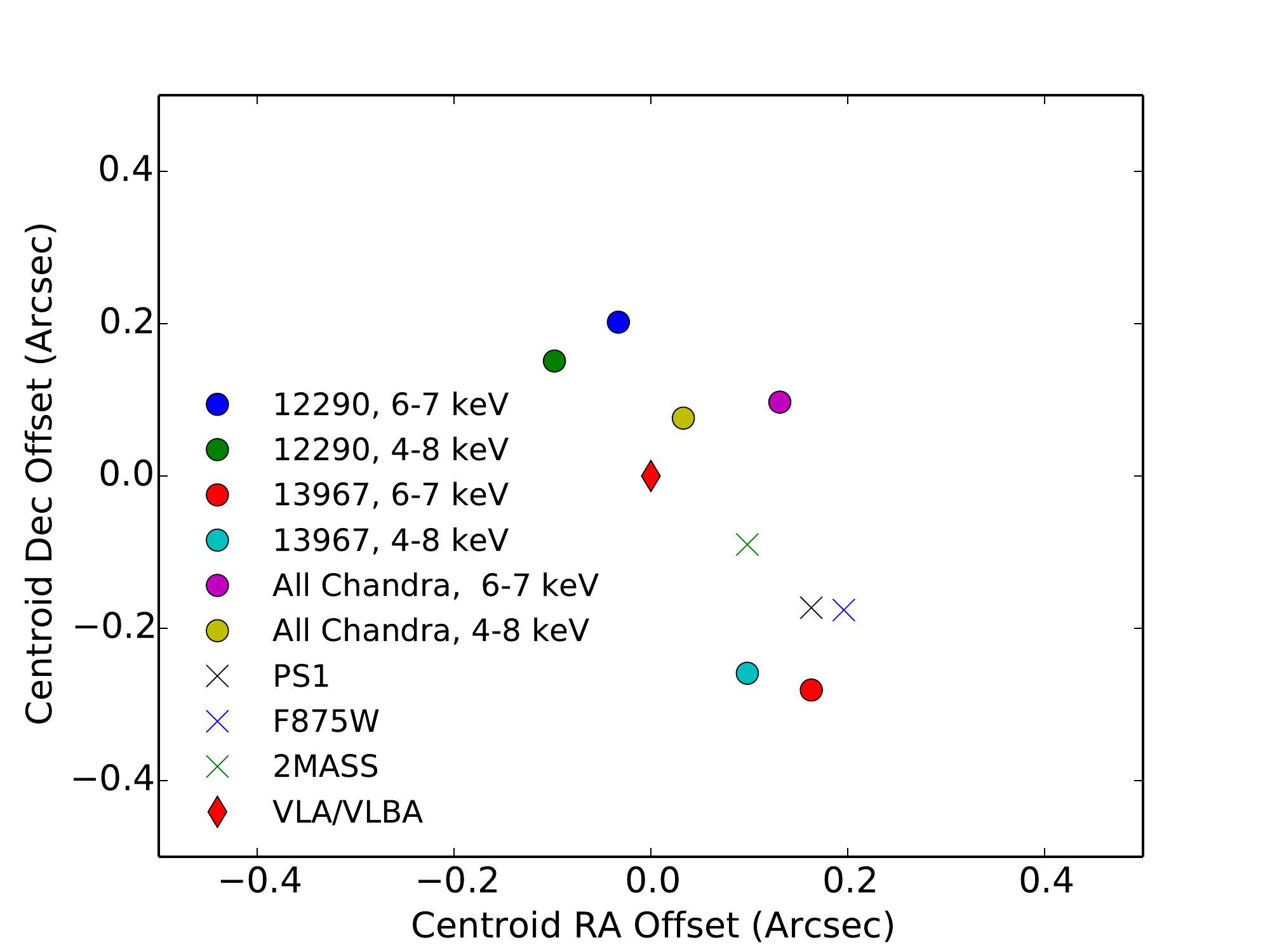}
\caption{Offsets in position of the galaxy nucleus in the X-ray, optical, and NIR as relative to radio position.  We find no significant offset in any of the observations, but very small offsets in absolute astrometry ($<$0.2$\arcsec$), suggesting the emission is coming from a single AGN down to these scales (50 pc).  The largest scatter is from the X-ray centroids, but a combination of all the X-ray observations to date finds a smaller scatter from the radio position.  Additionally, when combining optical with X-ray imaging of offset sources, the X-ray observations show reduced offsets (e.g. 12290 $<$0.1$\arcsec$).  The optical (PS1) and 2MASS images also have offsets in absolute astrometry, but these can be reduced to $<0.05\arcsec$ by combining imaging of offset stars.   }
\label{astrom}
\end{figure}

	  In a past study \citep{Fabbiano:2011:431}, the optical \hst image and water maser seen in the radio were found to be offset from each other, suggesting the presence of two AGN.  We therefore compare the centroids of all the radio, optical, NIR, and X-ray imaging (Figure~\ref{astrom}).  The absolute position of the dynamical center was estimated to within 1 mas based on the mean position of the low-velocity maser features \citep[R.A.$_{\mathrm{BH}}=162.09777^{\circ}$, decl.$_{\mathrm{BH}}=-25.162077^{\circ}$;][]{Kondratko:2008:87} and we use this observation as the absolute reference.  We find the hard X-ray, UV, optical, NIR, and UV emission are all coming from a single point source within a radius of $<0.2\arcsec$ (1$\sigma$).  This is fully consistent with astrometric errors in the X-rays of $\pm0.3\arcsec$ (1$\sigma$) based on Sgr A* field and $\pm0.15\arcsec$ (1$\sigma$) in the Hubble Deep Field  as well as well calibrated sky surveys such as 2MASS (0.3$\arcsec$ accuracy for extended sources like NGC 3393 at 1$\sigma$)\footnote{For details on the accuracy of astrometry with \Chrsh, see {\tt http://cxc.harvard.edu/cal/ASPECT/improve\_astrometry.html}.}.
	  
\subsubsection{Testing Chandra Imaging for a Dual AGN}

 We next test for the significance of the two reported nuclei using modeling and other techniques.  The 0.5$\arcsec$ pixel size of \Chr ACIS provides a limit to the resolution at which structures can be unambiguously detected.  Super-resolution can be achieved using the high cadence aspect solution and binning sky pixels at sub-integer sizes which will, on average, give a good approximation to the actual arrival direction of the photon. However, the small separation of the two proposed nuclei in past studies (0.6\arcsec) is very close to the telescope resolution (0.6$\arcsec$, 90\% confidence limits using ACIS-S).  Also, there are known PSF asymmetries that appear at small  separations ($<$0.8\arcsec).  Additionally, the previously claimed detection of a dual AGN was based on the 6-7~keV emission, where the collecting area of \Chr is low and the PSF is somewhat worse.   Finally, the previously reported detection combined two images and a possible incorrect offset between the two images would appear as two distinct sources\footnote{For more details on the \Chr PSF and its asymmetry, see {\tt http://cxc.harvard.edu/cal/Hrc/PSF/acis\_psf\_2010oct.html} and {\tt http://cxc.harvard.edu/ciao/caveats/psf\_artifact.html}.}.  Due to these complicating issues it is important to study the \Chr imaging in detail.

	 We first study the native \Chr PSF with EDSER.  To generate a simulated PSF, we use the \Chr ray-tracing program {\tt ChaRT} \citep{Carter:2003:477} and the {\tt MARX} software version 5 to project the ray-tracings onto the ACIS-S detector.  {\tt ChaRT} takes as inputs the position of the point source on the chip, the exposure time, and the spectrum of NGC 3393.   To create simulated images of a spatially offset dual AGN, we use the CIAO tool {\tt reproject\_events}.  The \Chr PSF was created by simulating a point source at the same position on the detector in each observation separately.  We run an on-axis ray trace simulation at 6.4 keV with 20000 counts using pixadj=EDSER and AspectBlur=0.19$\arcsec$, which represents the telescope pointing uncertainty. We find the \Chr PSF to have a FWHM of 0.8\arcsec when fit by a Moffat, Gaussian, or radial profile.  The radius of the 50\% energy enclosed function is 0.5$\arcsec$.
	
\begin{figure*} 
\centering
\subfigure{\includegraphics[width=8.2cm]{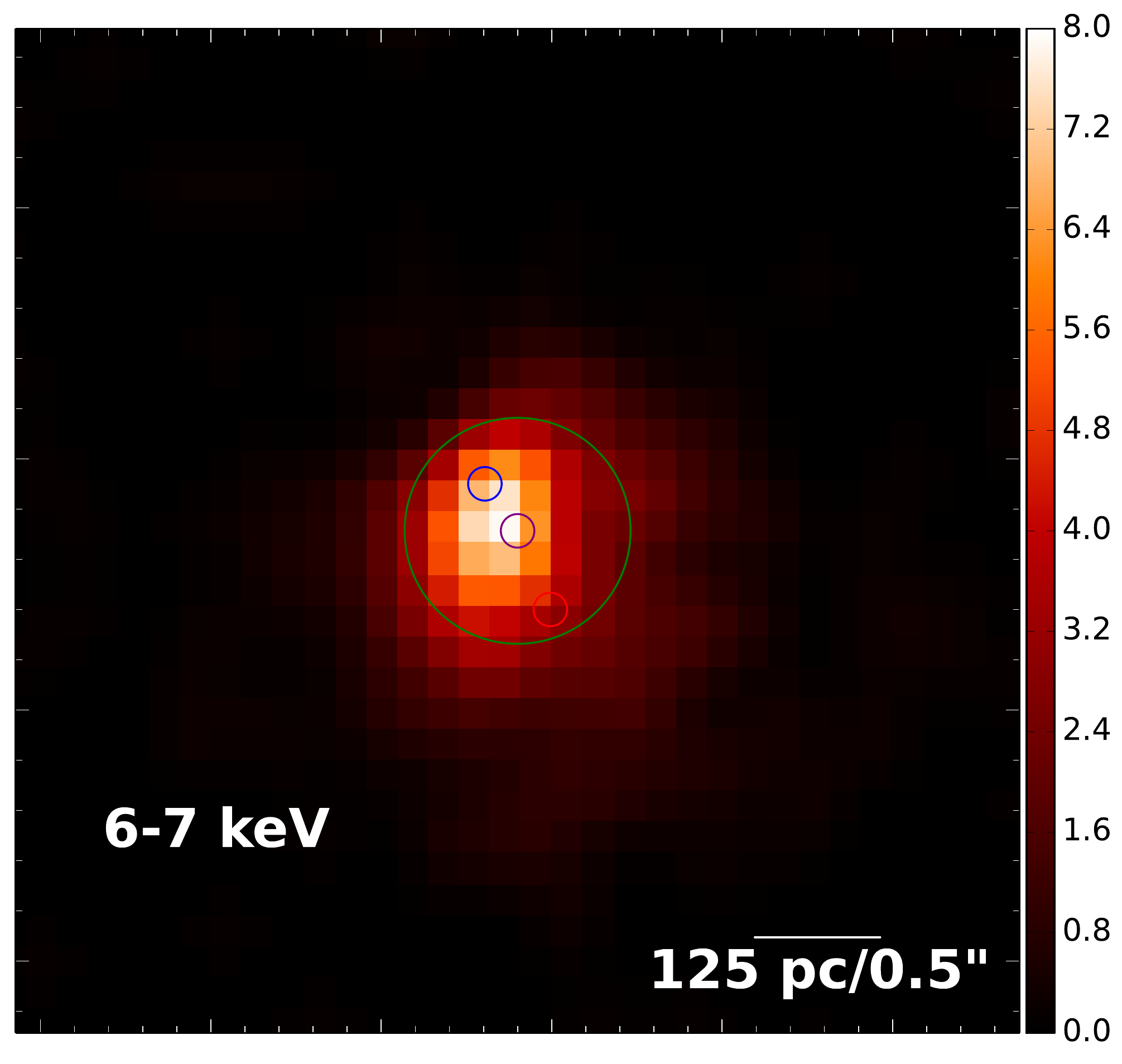}}
\subfigure{\includegraphics[width=8.2cm]{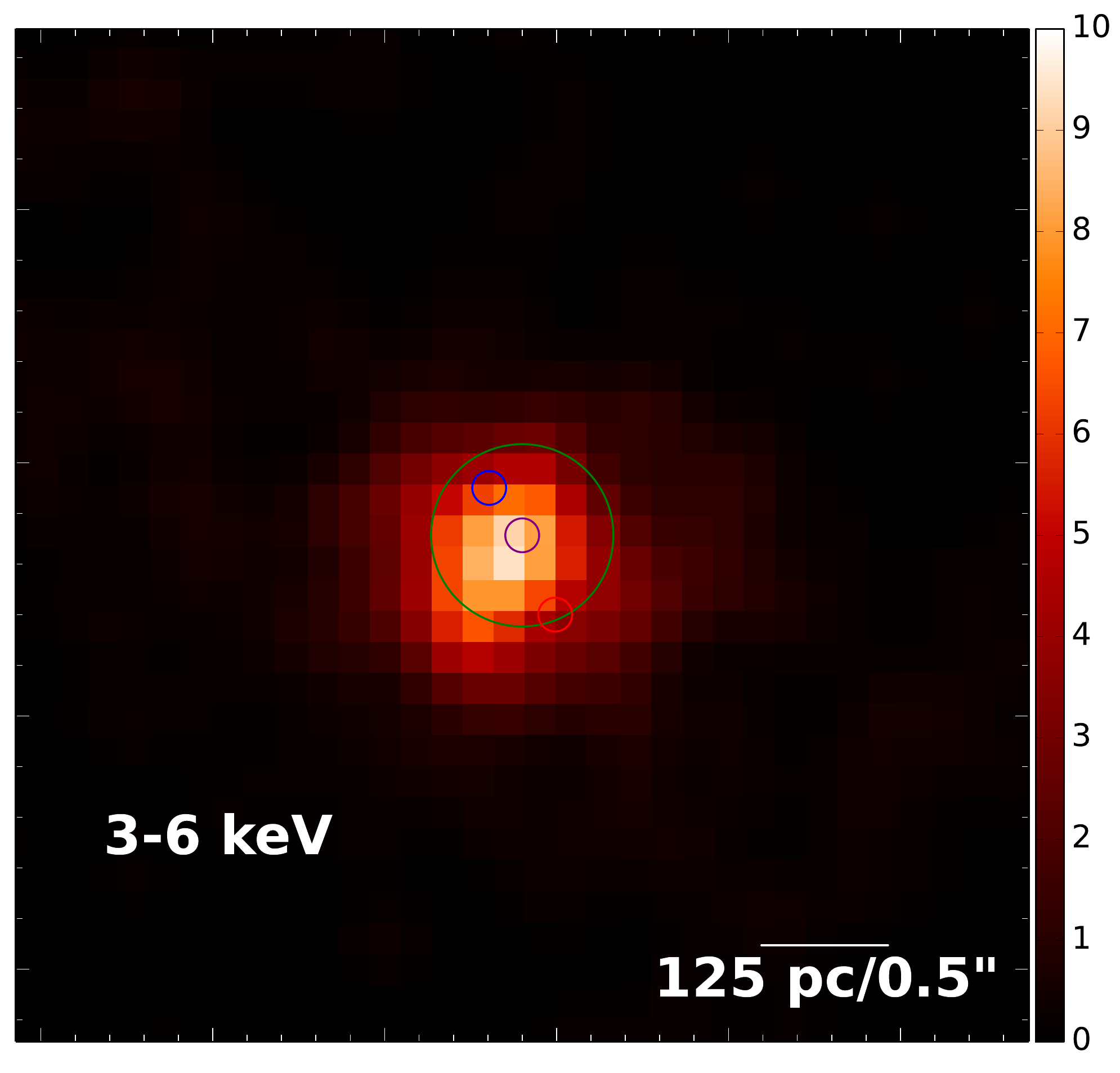}}
\vfill
\subfigure{\includegraphics[width=8.2cm]{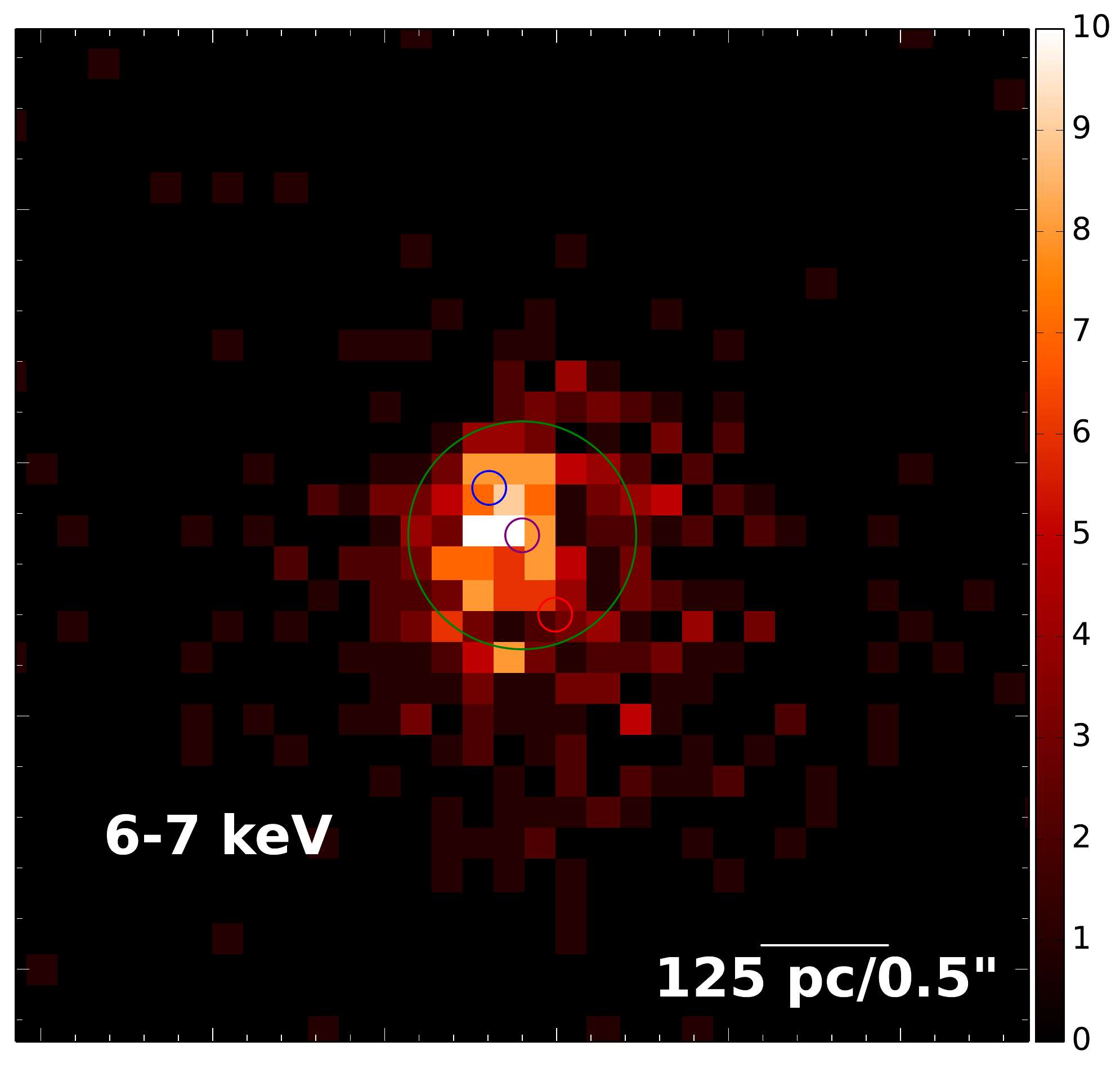}}
\subfigure{\includegraphics[width=8.2cm]{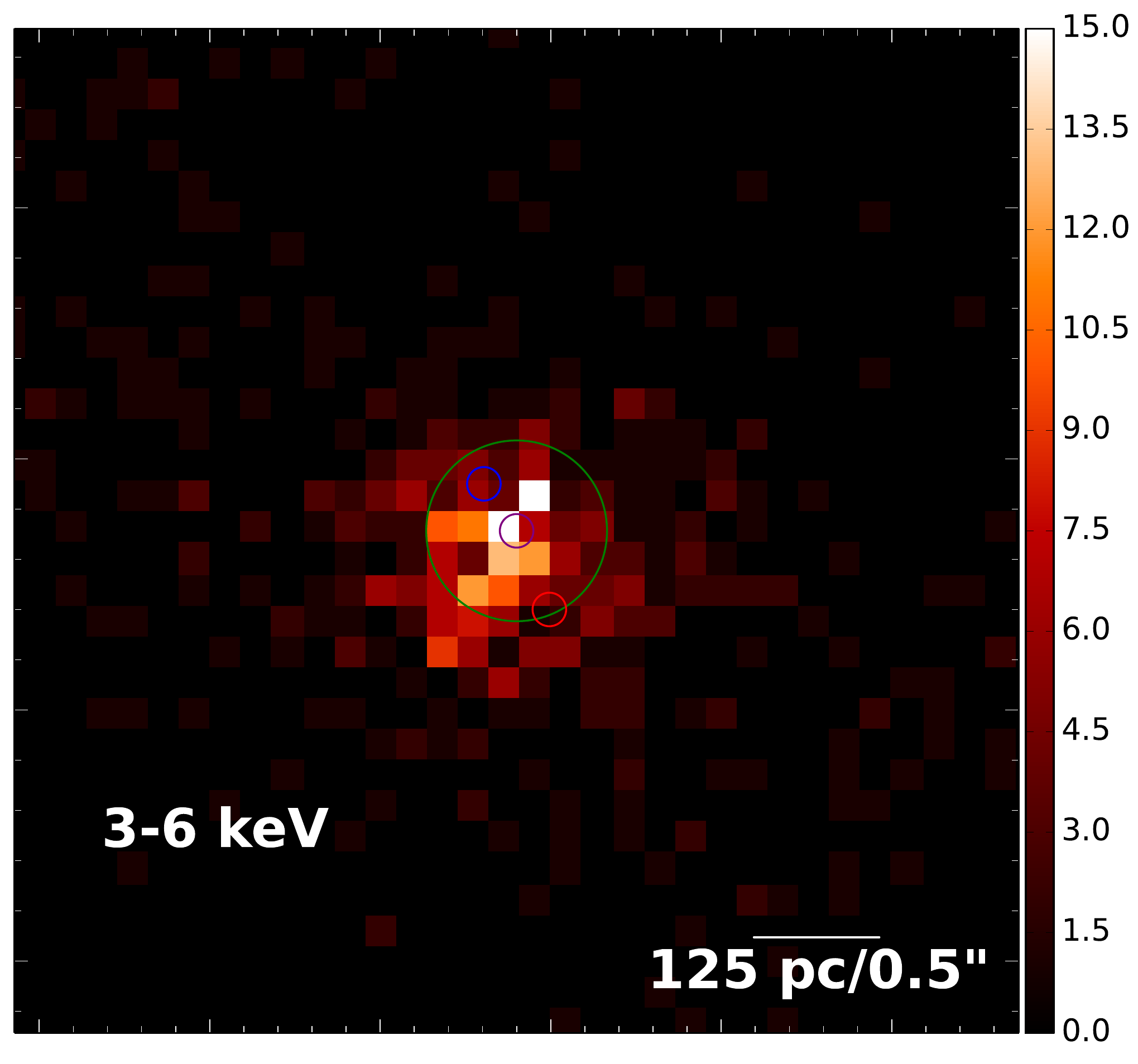}}
\caption{Stacked X-ray image with $3.1\times$ deeper imaging rebinned to 1/4 pixel (0.125\arcsec) resolution with 0.25$\arcsec$ smoothing (top) or no smoothing (bottom). Highlighted as circles are the 6-7~keV centroid (purple), PSF FWHM (green), as well as the positions of the reported dual AGN from in past studies (blue and red).  The red circle indicates the secondary AGN reported at 6-7~keV but not at 3-6~keV in past observations.  The 3-6~keV image is on a higher linear scale because of the larger number of counts.   We see no evidence of a dual AGN with the improved astrometry and deeper imaging, and the emission is consistent with the same single point source at both 6-7~keV and 3-6~keV.    }
\label{xraystack}
\end{figure*}

\begin{figure*} 
\centering
\includegraphics[width=14cm]{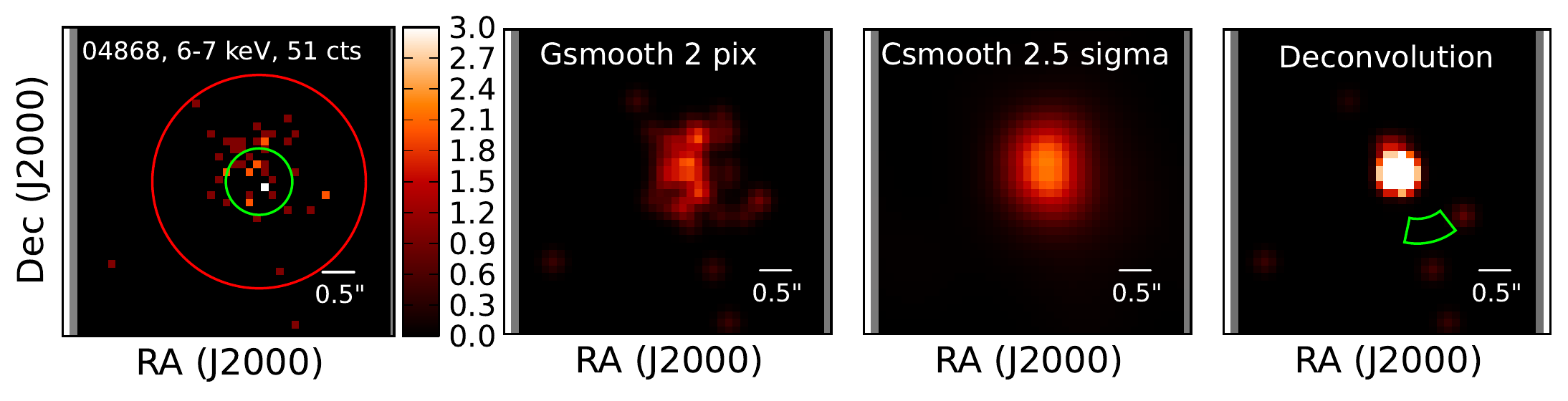}
\includegraphics[width=14cm]{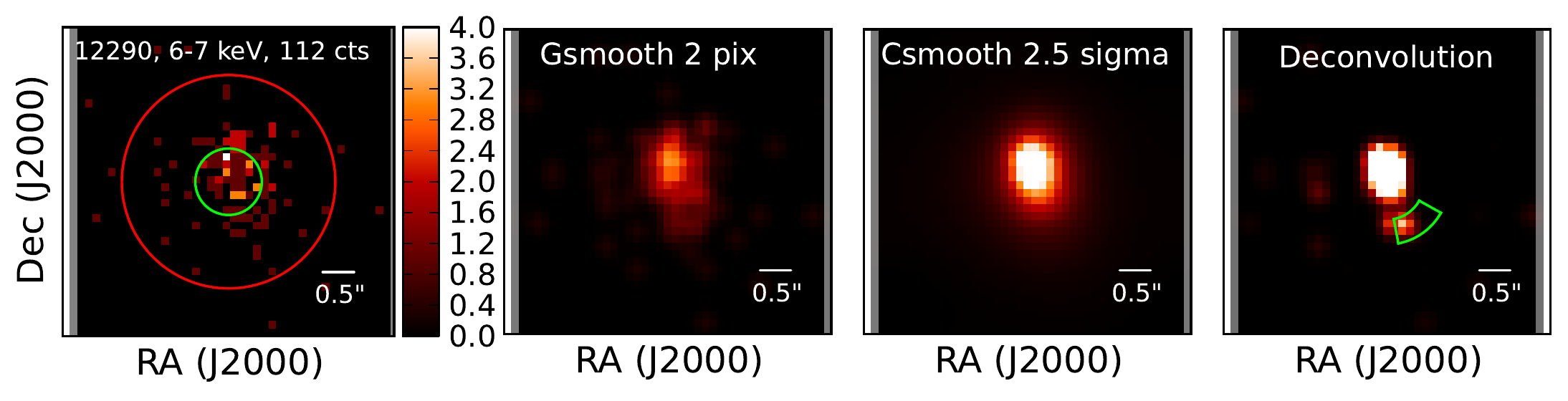}
\includegraphics[width=14cm]{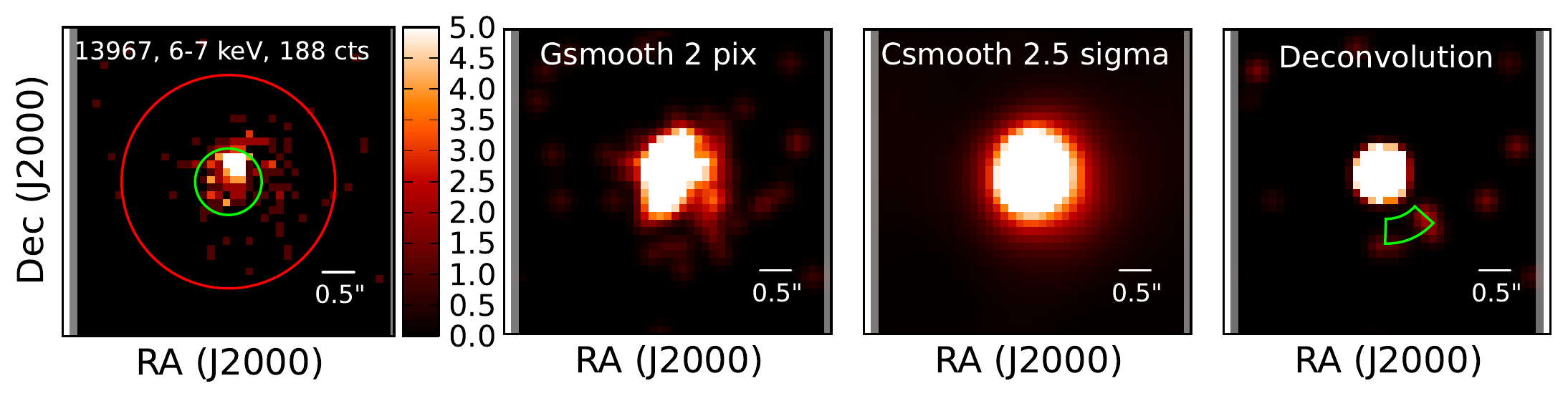}
\includegraphics[width=14cm]{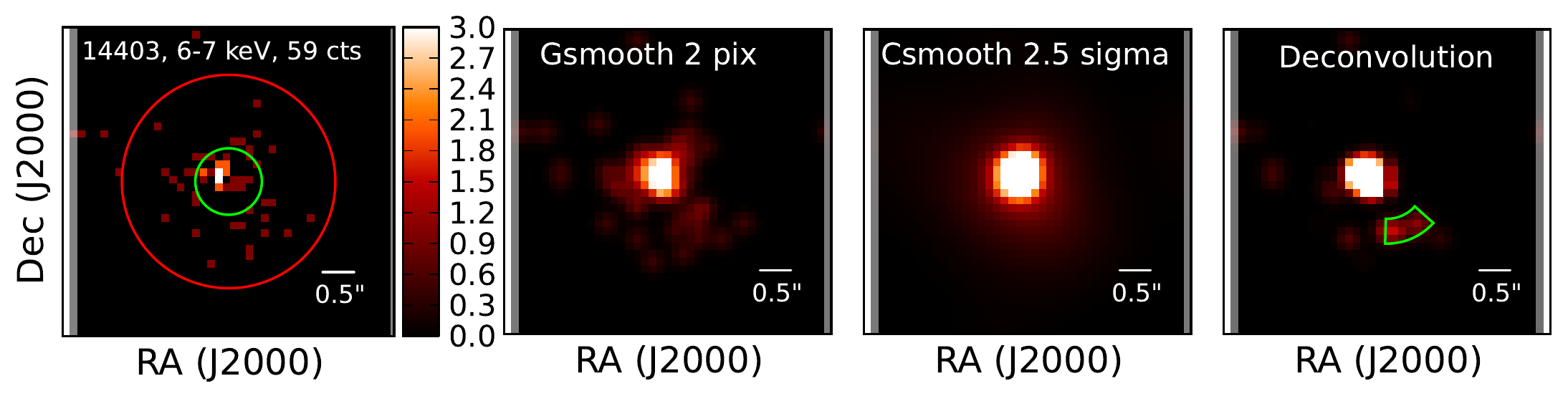}
\includegraphics[width=14cm]{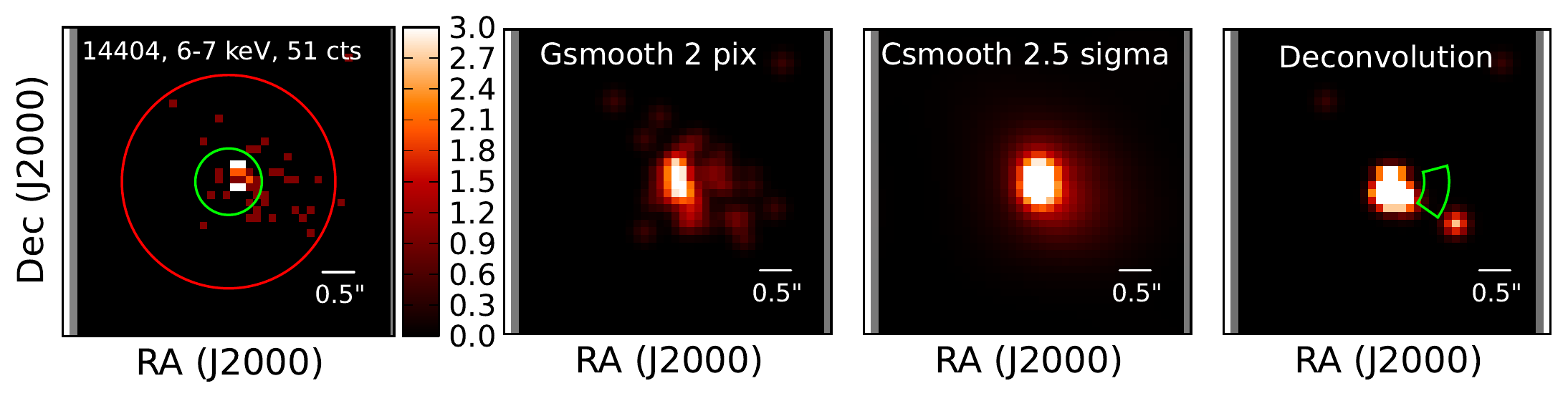}
\includegraphics[width=14cm]{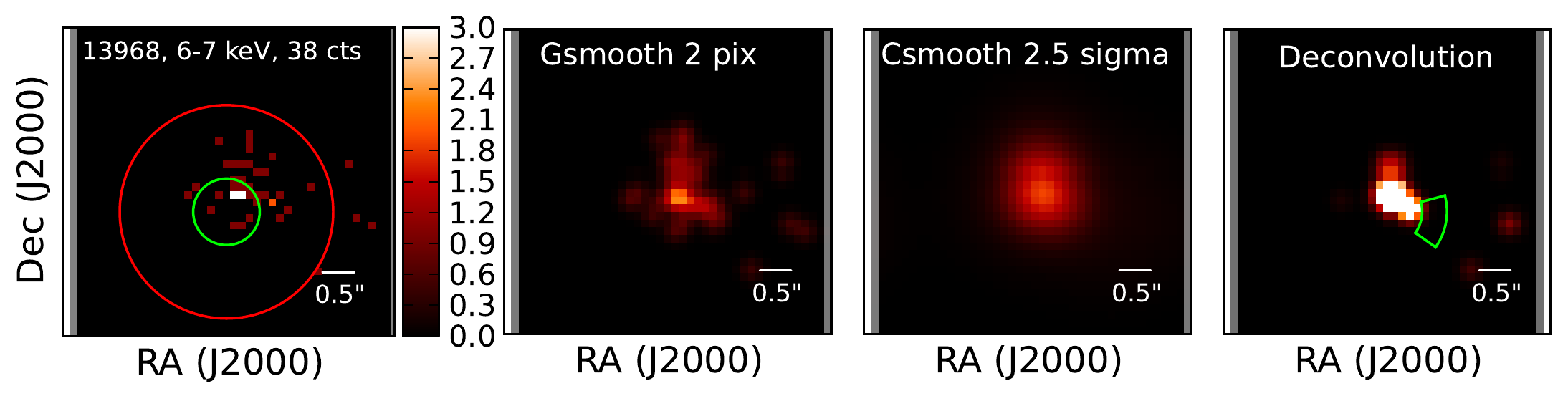}
\caption{From left to right:  data, gaussian smoothed, csmooth using Poisson statistic,  and deconvolved image in the 6-7~keV range at 1/4 sub pixel binning (0.125$\arcsec$) for different observations of NGC 3393.  A green and red circle indicate the 50\% and 90\% energy encircled function.    Note how the 3 pixel Gaussian smoothing shows features that are not present in the csmooth which accounts for Poisson statistics in the low count regime.  The green shape indicates the position of a hook that is a known HRMA PSF asymmetry found in brighter observations that depends on the roll angle of the spacecraft plotted with CIAO tool {\tt  make\_psf\_asymmetry\_region}.   We find that there is excess emission offset from the PSF by 0.8$\arcsec$ at 6-7~keV, but its orientation, separation, and brightness level ($<$10\% of central PSF or 1-8 counts) are consistent with the hook asymmetry feature of the HRMA found in brighter images.}
\label{all_deconv_67}
\end{figure*}

	To detect a dual AGN we look at the stacked as well as the individual images deconvolved by the expected PSF.  The stacked image from 3-6~keV and 6-7~keV can be found in Figure~\ref{xraystack}.  With more than a factor of three deeper \Chr imaging at 6-7~keV the summed imaged shows no evidence of a secondary AGN either at 6-7~keV or 3-6~keV.  The brightest source falls between the two previous detections and is consistent with a single AGN.  We also investigate all of the images separately at 6-7 keV and find no evidence of a dual AGN (Figure~\ref{all_deconv_67}).
	
\begin{figure*} 
\includegraphics[width=15cm]{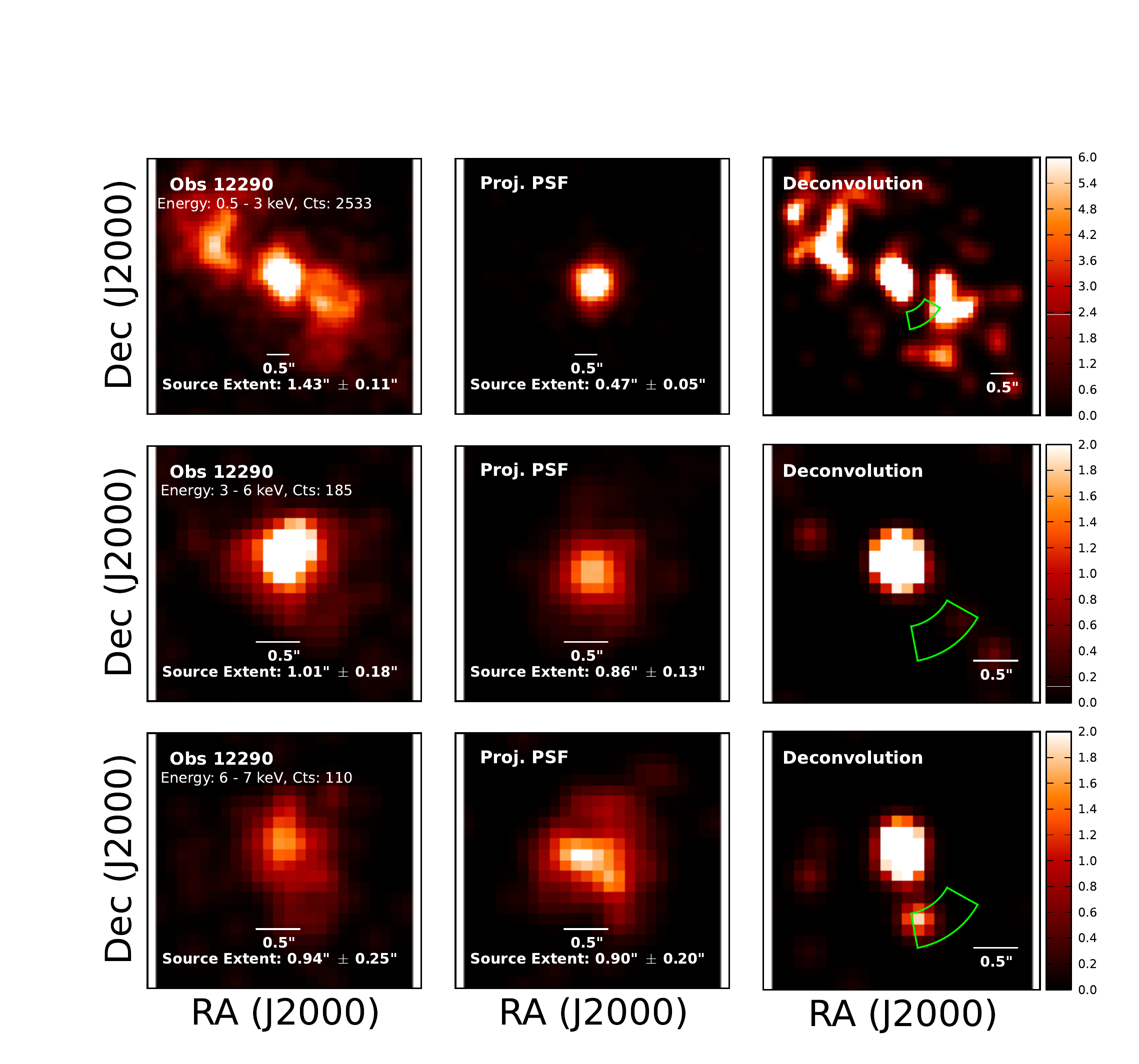}
\caption{{\it From left to right}:  data, projected PSF, and deconvolved image for obsid 12290 using the predicted PSF of NGC 3393.  The first row is 0.5-3~keV, the second row is 3-6~keV, and the bottom row is 6-7~keV.  We find extended emission in the 0.5-3~keV data, but no significant extended emission is found at 3-6~keV or at 6-7~keV.  The green shape indicates the position of a hook that is a known HRMA PSF asymmetry that depends on the roll angle of the spacecraft.     The extended emission below 3~keV is coincident with the \oiii and UV emission coming from the denser gas that is swept up by the leading edge of the radio jet.  The low counts (110 counts) limit the ability to find a dual AGN at scales below $\approx$1$\arcsec$.     }
\label{deconv}
\end{figure*}

	The deconvolved image for the deepest 70 ks exposure from the previously claimed dual AGN paper in different energy ranges can be found in Figure~\ref{deconv}. We find extended emission in the 0.5-3~keV data (source extent 1.43$\arcsec\pm0.11\arcsec$ with PSF source extent of 0.47$\arcsec\pm0.05\arcsec$), but no significant extended emission is found from 3-6~keV (1.01$\arcsec\pm0.18\arcsec$ for observation and 0.86$\arcsec\pm0.13\arcsec$ for the PSF)  and 6-7~keV (0.94$\arcsec\pm0.25\arcsec$ for nucleus and 0.90$\arcsec\pm0.25\arcsec$ for PSF).
	
\begin{figure*} 
\centering
\includegraphics[width=8.7cm]{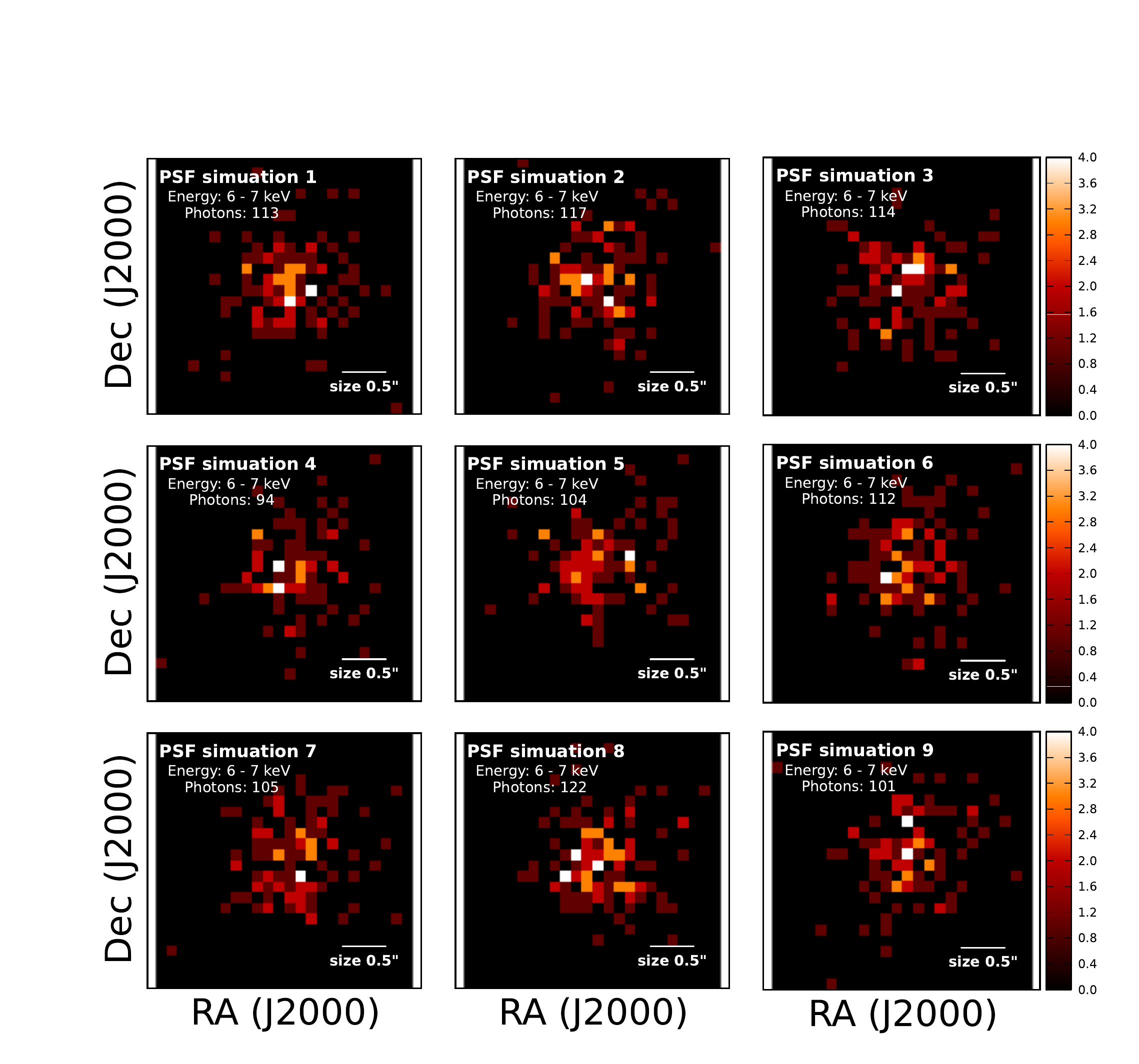}
\includegraphics[width=8.7cm]{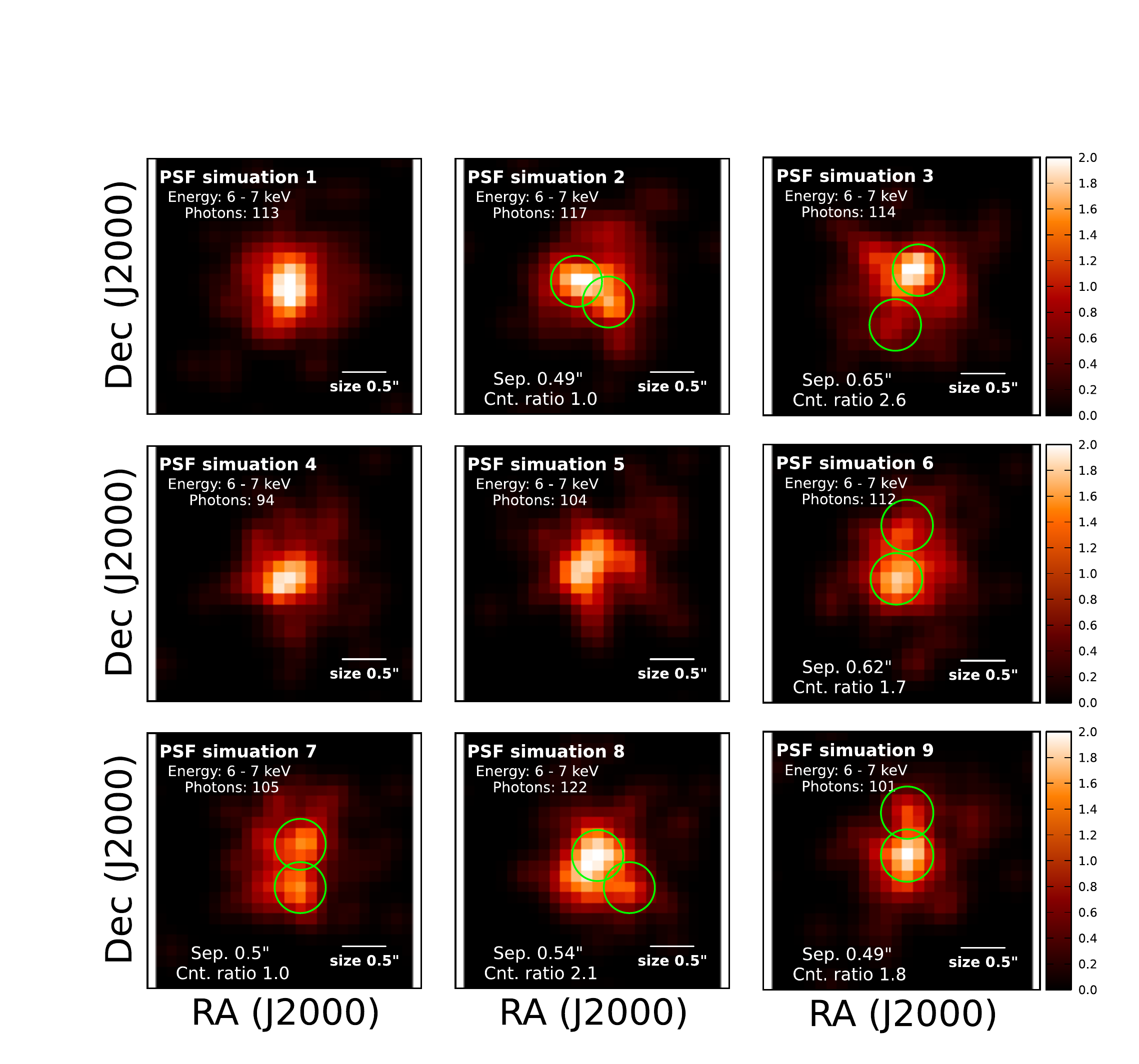}
\caption{Nine simulations of a single simulated \Chr PSF shown unsmoothed and smoothed.  We have used a Gaussian 2 pixel smooth kernel at 1/4 pixel binning following the sub-pixel binning and smoothing scales of in the purported dual AGN discovery paper \citep{Fabbiano:2011:431}.  Most of the images (6/9 or 67\%) show two bright peaks, despite being statistical noise from a single PSF.   The green circles are centered on the two brightest peaks in each image.   This is because of the low counts (1-3 per pixel) and that the smoothing kernel ($0.25\arcsec$) is significantly smaller than the resolution at 6-7 keV ($\approx0.8\arcsec$) causing Possion scatter noise to generate two distinct peaks.  We find that the average separation of the two fake peaks is $0.55\arcsec\pm0.07\arcsec$ with the secondary having on average $39\pm10\%$ of the total counts which is similar to the purported dual detection in NGC 3393 in separation (0.6\arcsec) and the secondary counts ($32\%$ of total), suggesting the detection is statistical noise.  }
\label{psf_fake}
\end{figure*}

	We then studied whether the purported dual AGN could be generated by Gaussian statistical noise in a single PSF (Figure~\ref{psf_fake}) with the same exposure as the 70 ks detection.  We ran nine simulations of a single simulated \Chr PSF.  We used a Gaussian 2 pixel smooth kernel at 1/4 pixel binning following the sub-pixel binning and smoothing scales of the purported dual AGN discovery paper \citep{Fabbiano:2011:431}.   We inspect the smoothed \chandra image and locate the two brightest peaks, when there is an increase in the smoothed counts compared to adjacent pixels.    {\bf We only include bright secondary peaks where the sum of the counts in a 0.3$\arcsec$ radius circle is greater than 25\% of the sum of the brightest peak.}  Most of the images (6/9 or 67\%) show two bright peaks at random orientations despite being statistical noise from a single modeled PSF.  This is because of the low counts (1-3 per pixel) and the fact that the smoothing kernel ($0.25\arcsec$) is significantly smaller than the resolution at 6.4 keV (0.8$\arcsec$), causing Possion noise to appear as two distinct peaks.  We find that the average separation of the two spurious bright peaks is $0.55\arcsec\pm0.07\arcsec$, with the secondary having on average $39\pm9\%$ of the total counts.  This is similar to the claimed dual detection in NGC 3393 in separation (0.6\arcsec) and in the secondary counts ($32\%$ of total), suggesting the detection is statistical noise. 

\subsection{Broadband Observations of a Compton-thick AGN}
\subsubsection{High-Resolution Chandra Imaging and Spectroscopy}

\begin{figure} 
\centering
\plotone{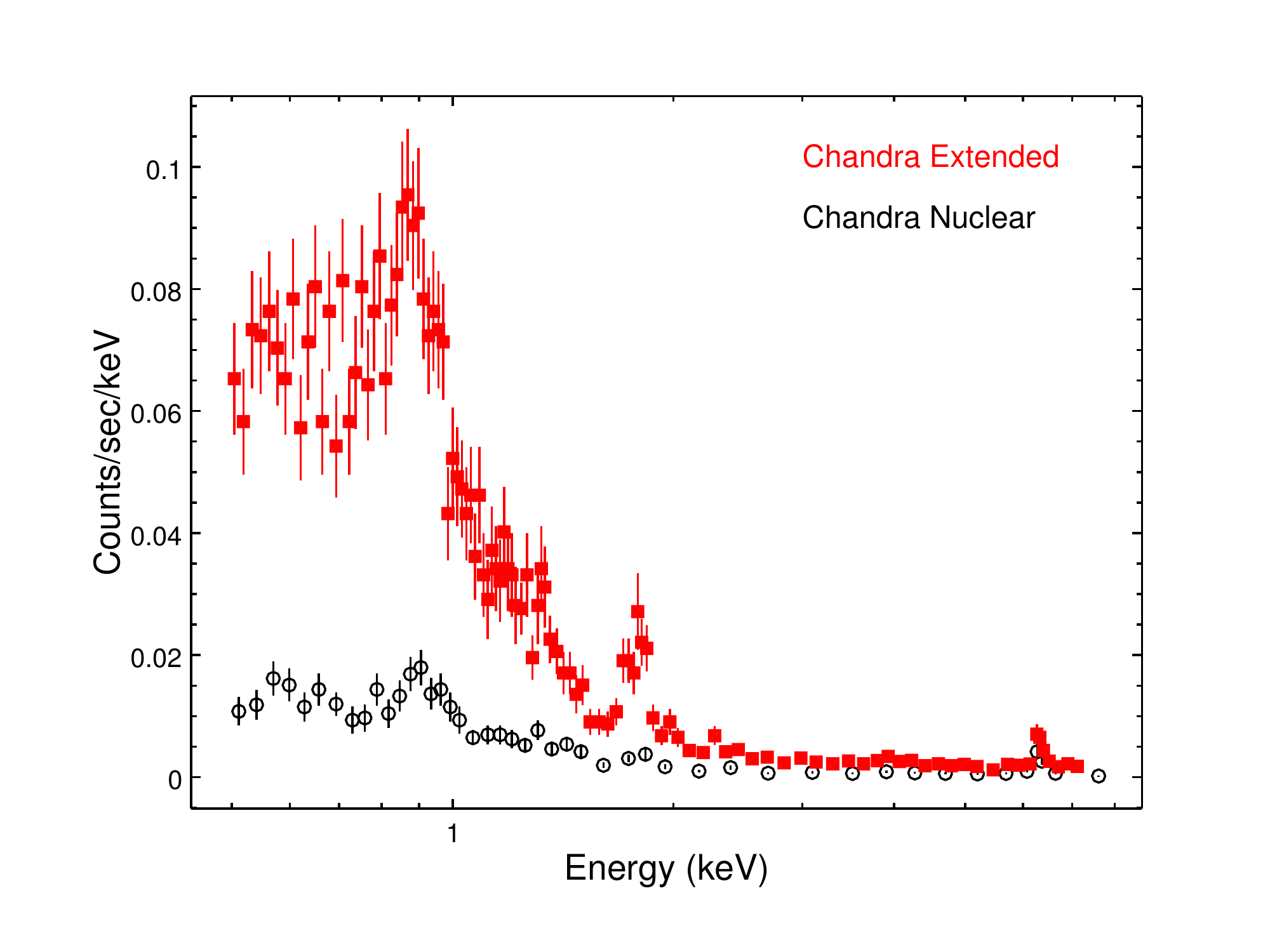}
\caption{Count rate from a 40$\arcsec$ aperture in \Chr (red filled squares), the same as the aperture used for \nustar and {\it XMM-Newton}, compared to the central 2$\arcsec$ (black circles).  Below 3~keV the majority of the emission is extended, with a significant peak at 1.8~keV suggesting significant photoionized emission.  In the 3-6~keV energy range, a smaller amount of emission is extended beyond the PSF emission (44\%). In the 6-8~keV energy range, there is no evidence that any of the emission is extended.  Because of this contamination, we focus on the $>3$~keV emission for measurement of AGN emission models.   }
\label{chandra_ext}
\end{figure}

We first examine the high resolution imaging from \Chr to determine the nature of soft extended emission in order to best model our \nustar data.  A plot of the nuclear PSF emission compared to a 40$\arcsec$ region can be found in Figure~\ref{chandra_ext}.  Below 3~keV the majority of emission is extended (81\%) based on comparing the background-subtracted count rates from a 40$\arcsec$ region ($0.056\pm0.003$~\cps) compared to a region enclosing the PSF ($0.011\pm0.001$~\cps).  In the 3-6~keV range, a significant amount of emission is still extended beyond the PSF emission (44\%; $0.0034\pm0.0008$~\cps~compared to $0.0019\pm0.0003$~\cps).  Between 6-8~keV, there is no evidence that any of the emission is extended  ($0.0019\pm0.0008$~\cps~compared to $0.0014\pm0.0003$~\cps), consistent with our 2D {\tt MARX} simulations.   A steep soft power law component with $\Gamma_s=3.3$ is a good fit to the extended \Chr data.\\

	The extended emission shows a significant peak at 1.8~keV which suggests strong photoionized emission.  We next focus on studying this emission at higher resolution using the grating spectroscopy from {\it Chandra}.  We note that the grating spectra are from a region of only 4$\arcsec$ in the cross-dispersion direction, so they correspond to a region similar in size to the nuclear region discussed in the last paragraph, rather than from the 40$\arcsec$ region where much of the extended emission below 3~keV is observed.   Using the other ACIS-S imaging data, we find that this 4$\arcsec$ region includes 53\% of the total background subtracted counts found in a 40$\arcsec$ region at 0.5-8 keV.    Gaussian line fitting was carried out using an automated procedure which fits the spectrum in 2$\AA$ intervals following \citet{Kallman:2014:121}.  The Gaussians are added to the model at wavelengths corresponding to known lines.  A summary of detected emission lines can be found in Table \ref{tab:chandra_lines}.  \\
	
	The most prominent observed spectral features are an emission line at 6.39~keV with an equivalent width of 1.11~keV associated with neutral iron, an emission line at 6.70~keV indicative of He-like iron with an equivalent width of 0.70~keV, and an emission line near 7~keV with an equivalent width of 1.13~keV which may be associated with a combination of the K absorption edge from near-neutral gas, emission from H-like iron (6.97~keV), and Fe K$\beta$ (7.06~keV).  The high value of the equivalent width of the iron lines is consistent with Compton-thick AGN \citep[e.g.][]{Krolik:1987:L5,Levenson:2002:L81}.

\begin{figure} 
\plotone{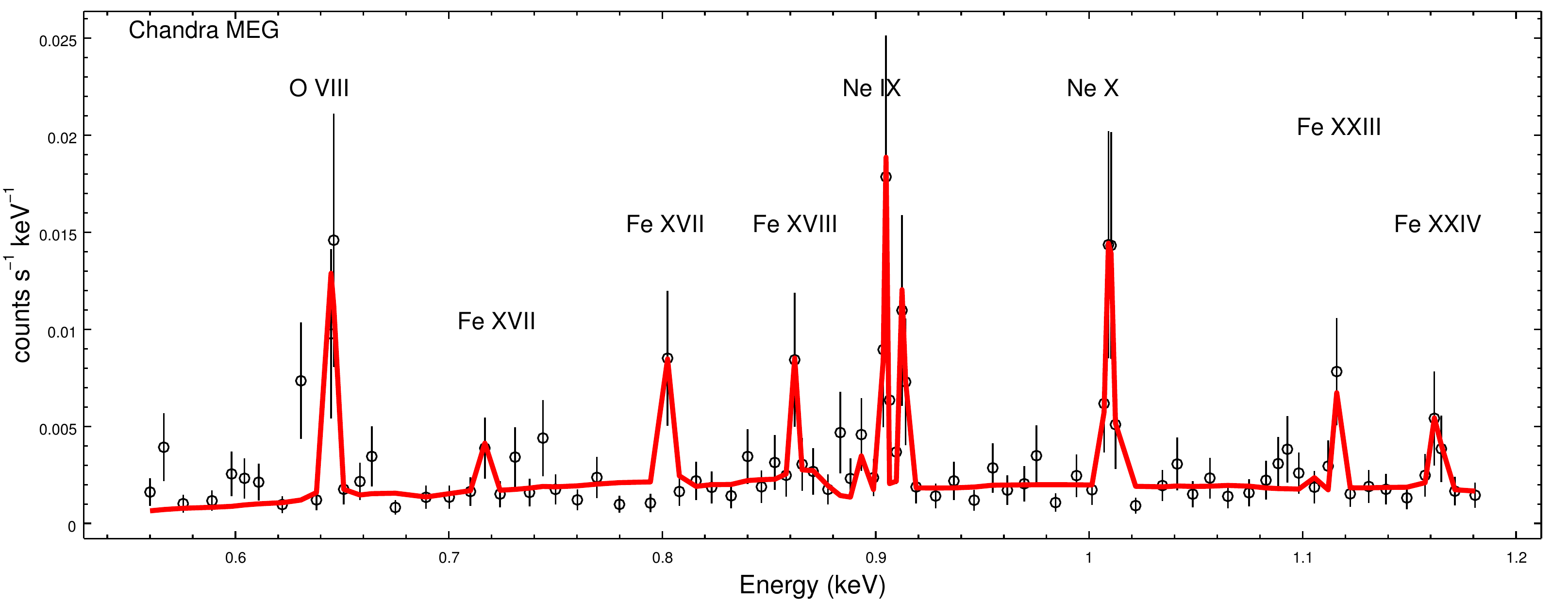}
\plotone{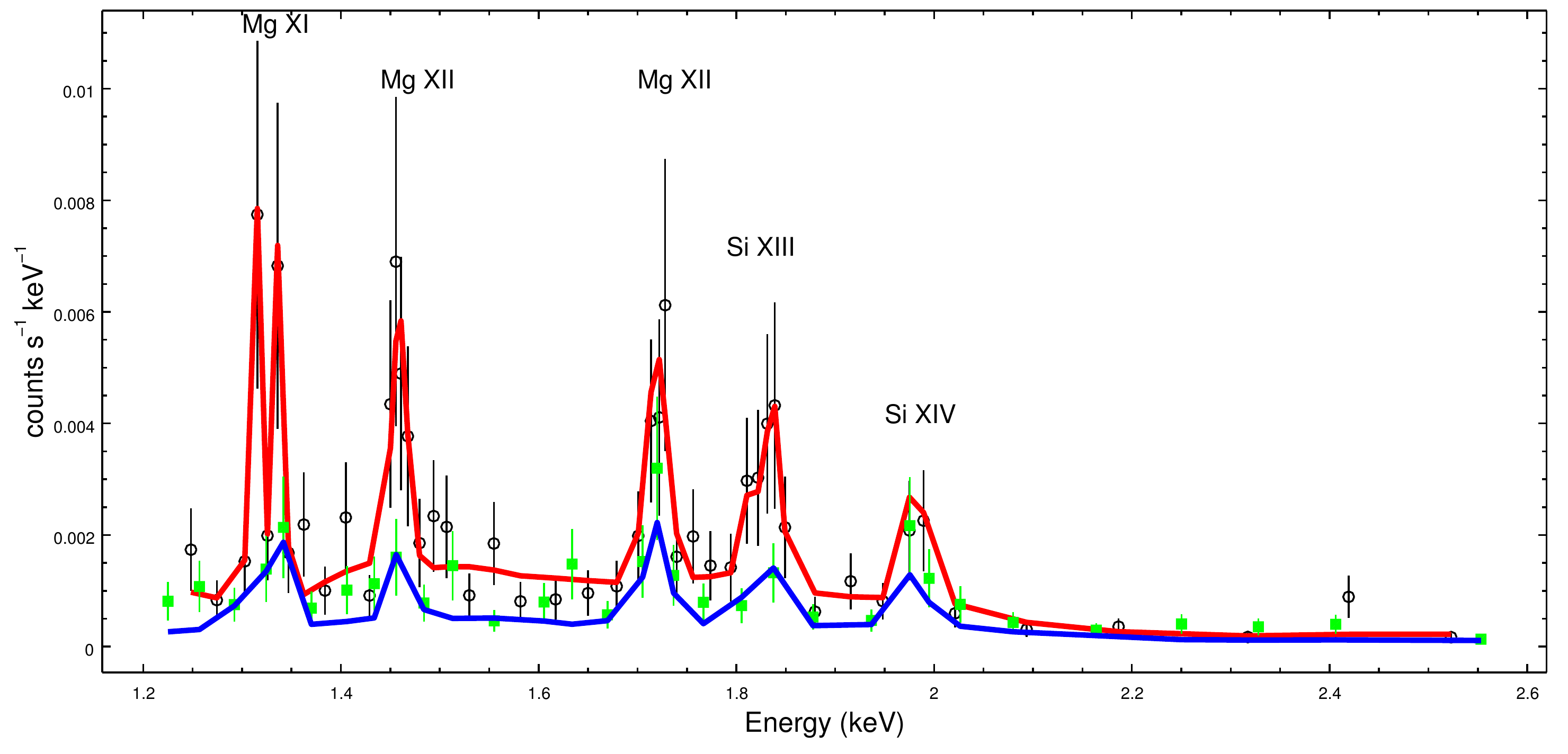}
\plotone{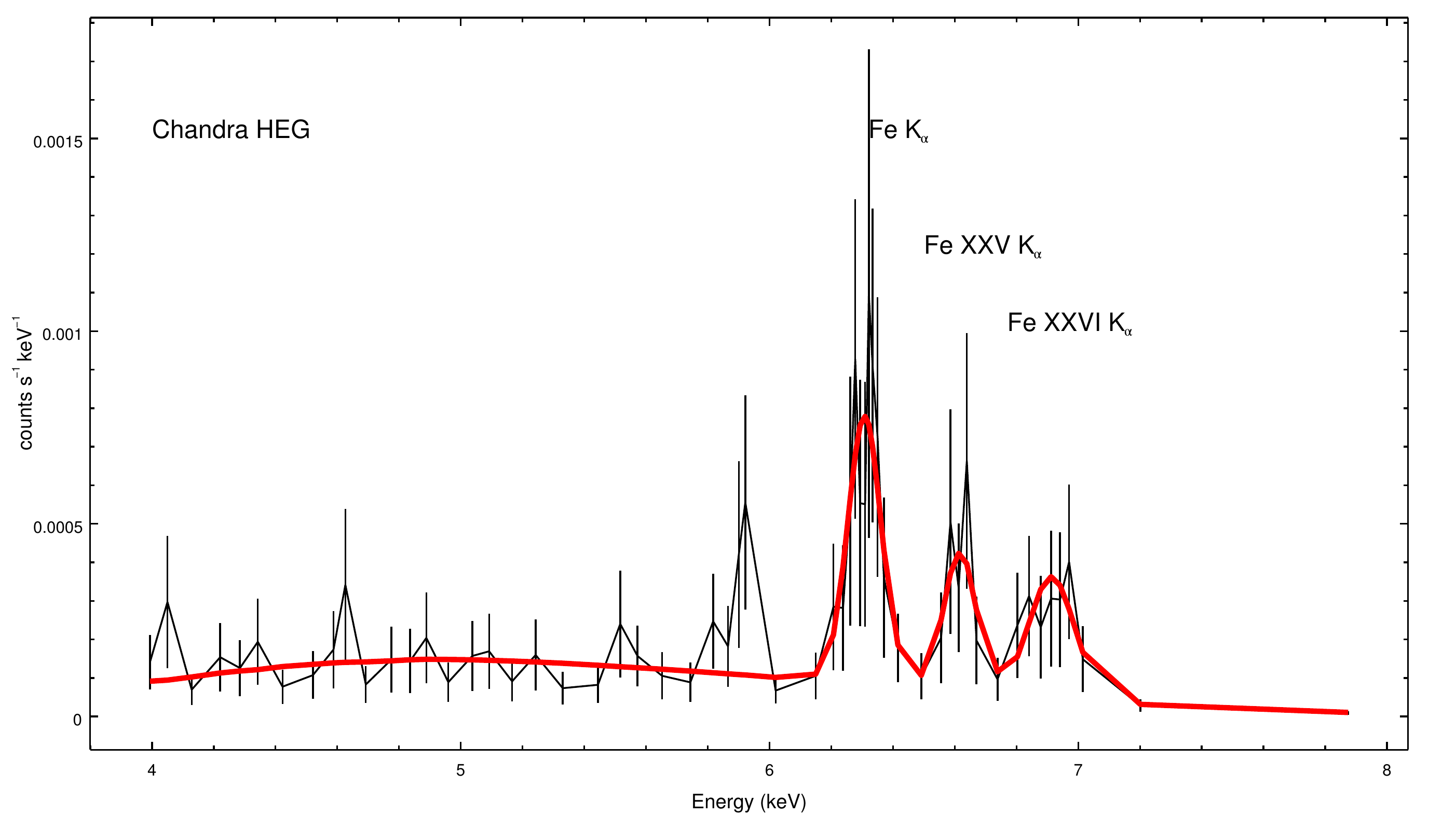}
\caption{Fits to 340 ks of \Chr grating data from the central 2$\arcsec$ aperture.  Even in the $2\arcsec$ nuclear region, the majority of the emission between 1 and 2~keV is from photionized line emission.  {\em Top:} MEG data (black data, red model) between 0.5-1.2~keV (10-22 \AA).   {\em Middle:}  Combined fit from \Chr grating spectra from MEG (black data, red model) and HEG (green data, blue model) between 1.2-2.5~keV (5-10 \AA).  {\em Bottom:}  Fit from \Chr grating spectra from HEG between 4.0-8.0~keV (1.5-3 \AA).  }
\label{dualfrac}
\end{figure}

\begin{table}[tbh]

\caption{\protect{NGC 3393 MEG and HETG Lines}} 
\label{tab:chandra_lines}
\begin{tabular} {llccr} 
\hline
Line & Observed & Lab & $\sigma$& Norm\\
& (keV) & (keV) &  (keV) & (1E-07)\\
    \hline
O VIII	&	0.653$\pm$0.002	&	0.654	&	0.0010	&	102.0	\\
Fe XVII	&	0.727$\pm$0.004	&	0.720	&	0.0003	&	16.5	\\
Fe XVII	&	0.813$\pm$0.004	&	0.826	&	0.0007	&	13.9	\\
Fe XVIII	&	0.873$\pm$0.003	&	0.873	&	0.0009	&	10.0	\\
Ni XIX	&	0.884$\pm$0.003	&	0.883	&	0.0003	&	5.9	\\
Ne IX	&	0.906$\pm$0.003	&	0.905	&	0.0005	&	6.5	\\
Fe XIX	&	0.916$\pm$0.003	&	0.917	&	0.0004	&	16.6	\\
Fe XIX	&	0.924$\pm$0.003	&	0.922	&	0.0007	&	13.6	\\
Ne X	&	1.022$\pm$0.003	&	1.022	&	0.0015	&	16.8	\\
Fe XXIII	&	1.131$\pm$0.003	&	1.129	&	0.0001	&	3.5	\\
Fe XXIV	&	1.177$\pm$0.003	&	1.168	&	0.0017	&	3.6	\\
Mg XI	&	1.330$\pm$0.003	&	1.331	&	0.0003	&	8.5	\\
Mg XI	&	1.352$\pm$0.003	&	1.352	&	0.0039	&	11.9	\\
Mg XII	&	1.478$\pm$0.005	&	1.472	&	0.0074	&	7.3	\\
Mg XII	&	1.741$\pm$0.003	&	1.745	&	0.0103	&	8.6	\\
Si XIII	&	1.837$\pm$0.003	&	1.839	&	0.0061	&	2.2	\\
Si XIII	&	1.861$\pm$0.003	&	1.865	&	0.0078	&	5.5	\\
Si XIV	&	2.006$\pm$0.004	&	2.005	&	0.0060	&	4.9	\\
Fe K$_\alpha$	&	6.388$\pm$0.03	&	6.40	&	0.0541	&	43.4	\\
Fe XXV K$_\alpha$	&	6.701$\pm$0.04	&	6.70	&	0.0519	&	25.7	\\
Fe XXVI K$_\alpha$	&	7.004$\pm$0.05	&	6.96	&	0.0727	&	38.1	\\
    \hline 
   \end{tabular}

\end{table}

\subsubsection{NuSTAR Spectral Fitting}
We then examine the \nustar data (3-70~keV) alone.    The \nustar FPMA and FPMB cross-normalizations were allowed to vary independently in the fit and are consistent with the 5\% error from measured calibration sources.  A simple absorbed power-law model does not fit the data well ($\chi^2/\nu=121/72$), but reveals that the spectrum is highly absorbed ($N_{\rm H}\sim10^{24}$~\pcmsq) and hard (effective photon index $\sim1.2$) -- indicating Compton-thick absorption and a significant contribution from a scattered (reflected) component. As a next step we use the approximate phenomenological models {\tt plcabs} \citep{Yaqoob:1997:184} for the transmitted component, and {\tt pexrav} \citep{Magdziarz:1995:837} to model the reflected component. The latter is implemented so that it reproduces only the reflection, $R<0$ in \xspec terminology. We link the photon index of the two components and fix the inclination at $\cos i =0.45$ and the high-energy cut-off at 200~keV \citep[following recent cut-off measurements e.g.][]{Gilli:2007:79,Malizia:2014:L25,Ballantyne:2014:2845}. In order to account for the clear line-like excess around the energy of the neutral iron K$\alpha$ line at 6.4 as well as the other iron emission lines at 6.7 and 7.0~keV we add a Gaussian component ($\sigma=10^{-3}$~keV) fixed at those energies. The best fit using this model ($\chi^2/\nu=51/70$) is found for $\Gamma=1.9\pm0.3$ and $N_{\rm H}=(3.1\pm0.6)\times10^{24}$~\pcmsq.

The model components used in the previous paragraph are only approximate; with the quality of the \nustar hard X-ray data we are able to also test more physically motivated models of the Compton-thick obscuring torus. One such model is \mytorus \citep{Murphy:2009:1549}. It includes both the transmitted and the scattered components, as well as fluorescent line emission computed for a torus with a half-opening angle of 60$^{\circ}$ through Monte Carlo simulations. A steep soft power law component with $\Gamma_s=3.3$ is included in the model in order to account for a part of the complex soft emission, as \mytorus predicts negligible flux below 5~keV for any line of sight through a Compton-thick torus. The best fit for the \nustar data is found for the \mytorus model in the coupled mode\footnote{This is the default mode, in which the components of the model are linked so that they represent reprocessing of radiation by a uniform toroidal distribution of gas. For details on the different usage modes of the \mytorus model and their interpretation we refer the reader to \citet{Yaqoob:2012:3360}.} with nearly edge-on inclination. A formally good fit ($\chi^2/\nu<1$) is found for fixed $\Gamma=1.9$ and any inclination greater than 70$^{\circ}$. Fixing the inclination at $i=80^{\circ}$ we find $N_{\rm H}=(2.5\pm0.2)\times10^{24}$~\pcmsq for fixed $\Gamma=1.9$, and $N_{\rm H}=(2.1\pm0.5)\times10^{24}$~\pcmsq for the fitted $\Gamma=1.7\pm0.2$.

\begin{figure} 
\centering
\plotone{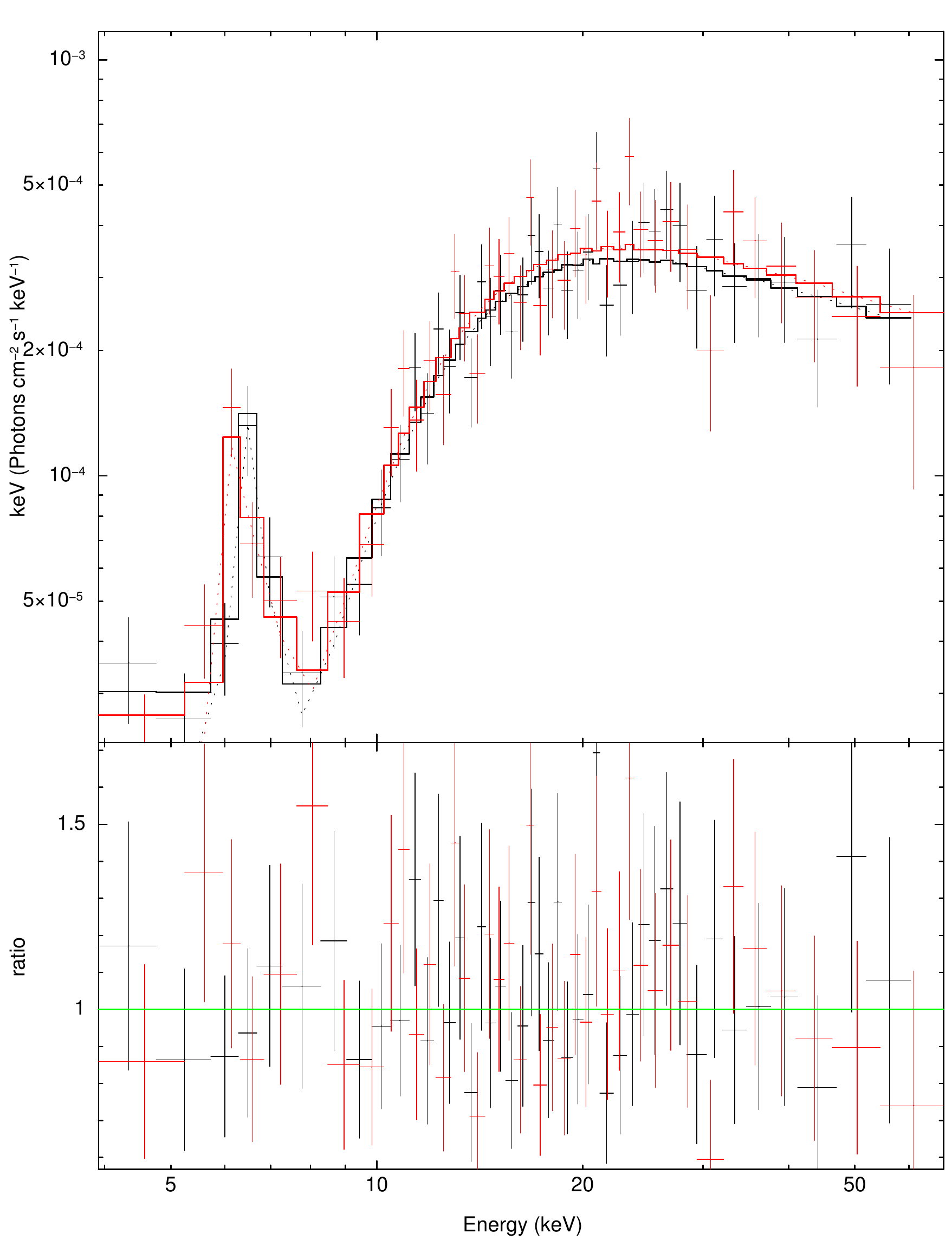}
\caption{Fit to \nustar data for NGC 3393 with \bntorus model. We fix the inclination to nearly edge-on, at $i=87^{\circ}$ and find $\Gamma=1.8\pm0.2$, $N_{\rm H}=(2.2\pm0.4)\times10^{24}$~\pcmsq, and the half-opening angle of the torus, $\theta_{\rm tor}=79_{-19}^{+1}$~degrees.   Similar properties within error are found from the \mytorus model as well.     }
\label{torus_nustar}
\end{figure}

Another physically motivated model for a torus, although the geometry is not exactly toroidal in this case, is the \bntorus \citep[]{Brightman:2011:1206} shown in Figure~\ref{torus_nustar}. We again fix the inclination to nearly edge-on, at $i=87^{\circ}$.  We find a statistically good fit ($\chi^2/\nu=50/70$) for the \nustar data with $\Gamma=1.8\pm0.2$, $N_{\rm H}=(2.2\pm0.4)\times10^{24}$~\pcmsq, and a half-opening angle of the torus $\theta_{\rm tor}=79_{-19}^{+1}$~degrees.  The small upper limit is present because the spectrum changes rapidly at very wide opening angles \citep{Brightman:2011:1206}.  The model yields an unabsorbed luminosity of $L_{2-10 \: \mathrm{keV}}=2.6\pm0.3\times10^{43} \ergps$  compared to an observed $L_{2-10 \: \mathrm{keV}}=1.7\pm0.2\times10^{41} \ergps$.  The ratio of the unabsorbed to absorbed flux is 149.   

In summary, \nustar data are consistent with the picture conveyed by both the \mytorus and the \bntorus models: a torus with relatively small covering factor (wide half-opening angle) seen through its Compton-thick wall with $N_{\rm H}\approx2\times10^{24}$~\pcmsq. Both the transmitted and the scattered components are needed for a good fit with nearly equal contributions above $\sim$2~keV. We therefore conclude that the hard X-ray data clearly point towards an edge-on, borderline Compton-thick torus.

The origin and the correct physical model for the soft X-ray emission are difficult to disentangle, since contributions from photoionization by the AGN and star formation in the galaxy, as well as Thomson-scattered AGN continuum, contribute a different fraction on different spatial scales. Under the assumption that star formation alone is responsible for the infrared flux, the IRAS luminosity yields a low star formation rate of $\approx$4 $\Msun/yr$ on 30 kpc scales \citep{Kondratko:2008:87}, corresponding to approximately $L_{2-10 \: \mathrm{keV}}=10^{40}$ \ergps, given the relationship between star formation and X-ray emission \citep{Lehmer:2010:559}.  As this value is more than an order of magnitude below the observed $L_{2-10 \: \mathrm{keV}}=1.7\pm0.2\times10^{41} \ergps$, this suggests star formation is not an important contributor above 2 keV.  The exact contribution of scattering, photoionization, and transmitted continuum to the soft X-rays is difficult to constrain because of model degeneracies other than that the photoionization lines dominate the emission below 2 keV based on the $\Chr$ grating data. However, we find that different parameterizations for the soft part of the spectrum fit jointly between \nustar and \Chr lead to self-consistent measurements of the intrinsic photon index and the column density, which match the values found based on the \nustar data alone within their 90\% uncertainties.  

In summary, we firmly establish that the hard X-ray data support a Compton-thick torus seen edge-on. This geometry agrees well with the fact that a disk water mega-maser has been detected in NGC~3393, since its detection implies a nearly edge-on view of the masing disk \citep{Kondratko:2008:87}. As the torus and the masing disk may correspond to different physical scales, the hard X-ray modeling implies a high degree of alignment of the AGN sub-structures.

\subsubsection{Variability and Spectral Fitting with Additional X-ray Observatories}
\begin{figure*}
\begin{minipage}[c][11cm][t]{.5\textwidth}
  \vspace*{\fill}
  \centering
\includegraphics[width=7.5cm]{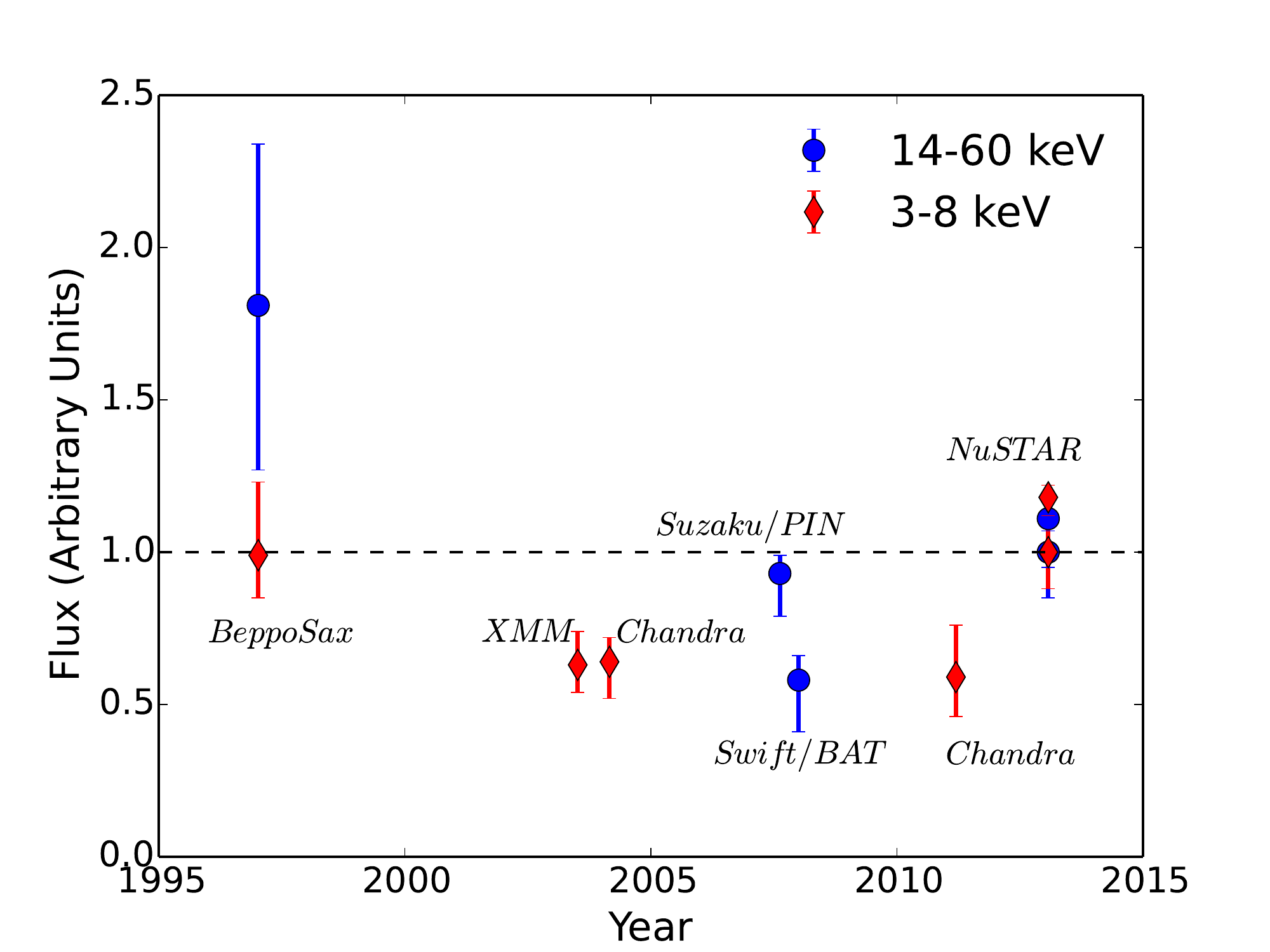}

  \label{fig:test2}\par\vfill
\includegraphics[width=7.5cm]{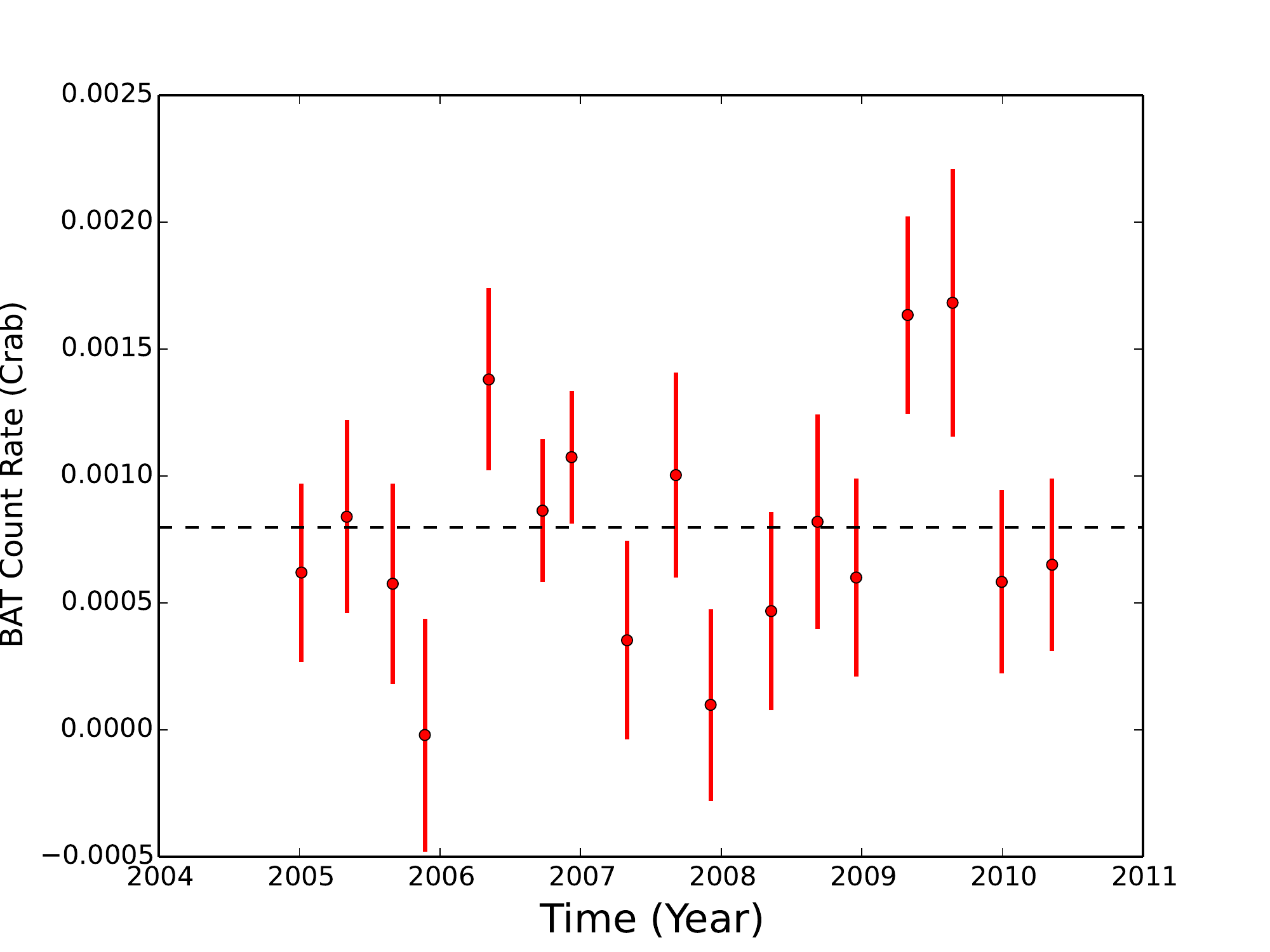}
  \label{fig:test3}
\end{minipage}
\begin{minipage}[c][11cm][t]{.5\textwidth}
  \vspace*{\fill}
  \centering
\includegraphics[width=8cm]{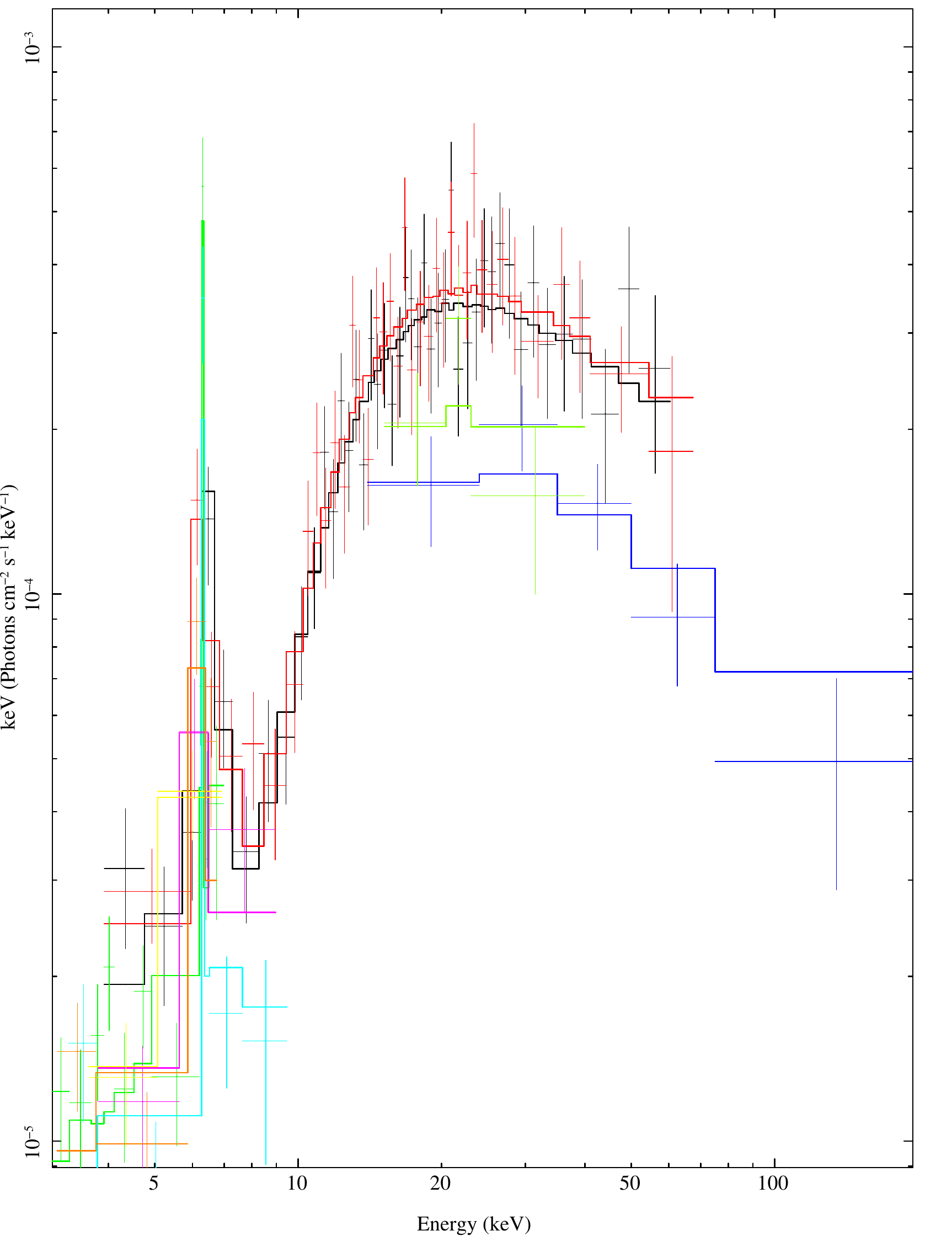}
  \label{fig:test1}
\end{minipage}
\caption{Tests for variability of NGC 3393 in the X-rays. {\em Upper Left}: Relative X-ray flux normalized to {\bf the \nustar 2013 observation}, from various X-ray observatories from 2004-2013 for NGC 3393, calculated by fitting the data with the normalization for each telescope allowed to vary.  The red points show normalized 3-8 keV flux while the blue points denote normalized 14-60 keV flux.   We find the data show no variability below 10~keV between 2004-2011, followed by an increase of $\approx$2 when the \nustar data was taken in 2013.   Above 10~keV, we find a similar increase of $\approx$2 when comparing the stacked BAT 2004-2010 observations and \nustarsh.  The lower sensitivity \suzaku PIN data shows no significant offset from \nustarsh.  {\em Bottom Left}: \swiftbat 70-month 14-195~keV light curve spanning 2004-2010 binned into 3 month intervals.  We do not see any evidence of variability based on fitting with a constant value.   {\em Right}: \bntorus model fit with the different data sets.  \nustar is shown in black (FPMA) and red (FPMB).  Above 10~keV, \swiftbat (dark blue) and \suzaku PN (light green) are shown.  Below 10~keV, \XMM PN (cyan), MOS 1 (orange), and MOS 2 (pink), as well as \Chr data from 2004 (green) and  2011 (pink) are shown.    }
\label{BAT_lightcurve}
\end{figure*}

	We consider whether NGC 3393 is variable, consistent with a transmission-dominated AGN, or constant, as expected for a reflection-dominated AGN.  We test variability first using similar telescopes to avoid cross-calibration issues.  We explored longer term variability in the hard X-ray band using the \swiftbat 70-month data taken between 2004 and 2010. We find the full 14-100~keV band light curve of a region centered on NGC 3393, binned in 3 month intervals, is consistent with a constant source model based on a chi-squared test (Figure~\ref{BAT_lightcurve}). We can also exclude variability at levels greater a factor of two when the data is binned into year long timescales at the 90\% level.  Splitting the \swiftbat data into several energy bands shows no significant variability ($>90\%$) either.  In the \Chr data taken in 2004 and 2011-2012 we also find no variability within the observation or between the observations at the 90\% level.  This is consistent with the findings of the \Chr Source Catalog which finds there is definitely no variability in the soft and medium energy bands ($<$2~keV) and likely no variability in the hard band based on a Gregory Loredo variability algorithm.   Finally, within the XMM observation we see no evidence of variability in the PN or MOS cameras at the 90\% level which agrees with the 3XMM-DR4 XMM catalog.  In summary there is no evidence of variability within any of the specific observations for NGC 3393.
	
	We finally fit all of the X-ray data using the \bntorus model and allow the normalization to float (Figure~\ref{BAT_lightcurve}).  The data seem consistent with a constant flux between 2003-2012, followed by a 3$\sigma$ increase in flux when the \nustar observation was taken in 2013.  {\it INTEGRAL} and \suzaku HXD have found \swiftbat cross-calibration to be systematically lower than their nominal values by the same factor of 0.82$\pm$0.03 factor at 14-100 and 14-70~keV, respectively, using different studies \citep{Molina:2013:1687}.  Given this systematic offset, the variability is only significant at a level of 1.8$\sigma$.  There seems to be a bright state during the 1997 \bepposax PDS observation, though it is difficult to exclude source contamination in this observation because of the nearby source in the BAT maps.  Further studies with the high sensitivity of \nustar would be necessary to confirm this high energy variability and to better understand the cross-calibration using very bright sources.
	
	We finally refit all of the X-ray data using the \bntorus model to confirm our conclusions about the torus based on the \nustar data alone.  We exclude the \bepposax data because of possible source contamination (see 2.3).  With an inclination to nearly edge-on, at $i=87^{\circ}$, we find a statistically good fit ($\chi^2/\nu=145/138$) with $\Gamma=1.82\pm0.09$, $N_{\rm H}=(2.34\pm0.19)\times10^{24}$~\pcmsq, and the half-opening angle of the torus, $\theta_{\rm tor}=78_{-3}^{+1}$~degrees.  These values are all consistent within the errors to the \nustar only fits.  The data is consistent with no high energy cutoff with a lower limit of $E_c>$169~keV when fitting a \pexrav and a high energy cutoff.
	
\subsubsection{ Intrinsic Luminosity, Eddington Ratio, and X-ray Radio Loudness}

We can calculate the expected bolometric luminosity ($L_{\mathrm Bol}$) of NGC 3393 using optical and MIR measurements  that can be compared to the X-rays.  This is critical because of the difficulty estimating the intrinsic luminosity based on the X-rays in Compton-thick AGN.  We calculate an extinction-corrected \oiii luminosity of of $1.5\times10^{41} \ergps$ from a 2.4$\arcsec$ diameter region or $3.9\times10^{41} \ergps$ from a 4.8$\arcsec$ diameter aperture  from the SNIFS IFU.   \citet{Lamastra:2009:73} reported the luminosity dependent bolometric correction factors.   At $L_{\mathrm {[OIII]}}=10^{40}-10^{42} \ergpssh$, the correction is 142, suggesting a bolometric luminosity of $2.1\times10^{43} \ergps$ for a 2.4$\arcsec$ diameter region or $5.5\times10^{43} \ergps$ for a 4.8$\arcsec$ diameter region which includes the extended \oiii emission.  We use the extracted low resolution \spitzer spectra from the Cornell Atlas of \spitzer IRS Sources \citep[CASSIS;][]{Lebouteiller:2011:8}, with a tapered column extraction.  We measure \nev and \oiv lines with luminosities of $9.2\times10^{40} \ergps$ and $5.7\times10^{41} \ergps$, respectively.  The \nev luminosity corresponds to a bolometric luminosity of $8.3\times10^{44} \ergps$.  Based on the unabsorbed 2-10~keV luminosity of $2.6 \times 10^{43} \ergps$, we use the bolometric correction of 30 from \citet{Vasudevan:2009:1124}.  We find a bolometric luminosity of $7.8\times10^{44} \ergpssh$, which is consistent with the \nev estimates. 

We modeled the spectral energy distribution (SED) of NGC 3393 using the \citet{Assef:2010:970} 0.03-30$\micron$ empirical AGN and galaxy templates to understand the strength of the AGN emission in the IR. Each SED is modeled as a best-fit, non-negative combination of an old stellar component, a starburst, plus an AGN component. Only the AGN component is fit for dust reddening \citep[for details, see][]{Assef:2008:286,Assef:2010:970}.  The reddening corrected 6 $\micron$ luminosity of the best-fit AGN component is $(1.5\pm0.2)\times10^{43}\ergpssh$.  The modeling outputs $L_{6\, \micron}$, the derived intrinsic luminosity of the AGN component at rest-frame 6 $\micron$, as well as the reddening of the AGN component, $E(B-V)_{\mathrm{AGN}}$.  For the typical gas-to-dust ratio observed by \citet{Maiolino:2001:28} for luminous AGN, the nuclear reddening values of $E(B-V)_{AGN}=12.6\pm6.2$ implies a gas column of $N_H\approx1\times10^{24}$ \cmsqsh, consistent with the Compton-thick AGN interpretation.  

Combined with the measured mass of the SMBH harbored by the AGN we can estimate the Eddington fraction, $L_{\mathrm Bol}/L_{\mathrm Edd}$, where $L_{\mathrm Edd}$ is the Eddington luminosity.  VLBI maps of nuclear water maser emission have been used to measure the BH mass to be ($3.1\pm0.2$)$\times10^7\Msun$ \citep{Kondratko:2008:87} with a corresponding Eddington luminosity of $4\times10^{45} \ergpssh$. The Eddington ratio is then $\approx$0.2 based on the \nev or intrinsic 2-10~keV emission, or $\approx$0.1 based on the \oiii data.  

Based on the intrinsic 2-10~keV X-ray luminosity and the nuclear 4.9 GHz emission from component A, we can estimate the X-ray radio loudness parameter, $R_{\mathrm X}=\nu L_{\nu}$(5 GHz)$/L_\mathrm{X}$.  Radio quiet objects are typically found at $\log R_{\mathrm X}<-4.5$ \citep{Terashima:2003:145}. We find a value of -6.28 for NGC 3393 consistent with a radio quiet rather than radio loud object.

\section{Summary and Discussion}
We study the Compton-thick AGN NGC 3393 using recently obtained $NuSTAR$, $Chandra$, VLBA, and Keck data combined with archival \hst and VLA data.  This nucleus has been claimed to harbor a dual AGN based on previous \Chr imaging. We find:

\begin{enumerate}
\renewcommand{\theenumi}{(\roman{enumi})}
\renewcommand{\labelenumi}{\theenumi}
\item Using data from the VLA and VLBA we find the radio structure is from an AGN with a two-sided jet rather than being due to a dual AGN (Section 3.1.1).

\item Compared to the milliarcsecond absolute astrometry provided by the VLBA the astrometry results suggest the hard X-ray, UV, optical, NIR, and radio core emission are all coming from a single point source within $<0.2\arcsec$ (1$\sigma$, Section 3.1.3).  We find no evidence of an offset between the radio and \hst imaging   in agreement with early studies of NGC 3393 \citep[e.g.,][]{Schmitt:2001:199}, but contrary to the past study suggesting a dual AGN \citep{Fabbiano:2011:431}.  AO imaging from NIRC2 in the $K^\prime$-(2.12$\,\micron$) band suggests only a single nucleus with no evidence for a secondary AGN at 0.1$\arcsec$ scales.  This is consistent with other AO studies in the NIR \citep{Imanishi:2014:106} and mid-IR \citep[$8-13\,\micron$,][]{Asmus:2014:1648}.

\item With more than a factor of three deeper imaging in the hard X-rays with \Chrsh, we find the the previously claimed dual AGN detection is most likely spurious resulting from the low number of X-ray counts ($<$160) at 6-7~keV and smoothing of data with a few counts per pixel on scales much smaller than the PSF (0.25$\arcsec$ vs. 0.8$\arcsec$ FWHM).  We show that statistical noise in a single \Chr PSF generates spurious dual peaks of the same separation ($0.55\pm0.07\arcsec$ vs. 0.6\arcsec) and flux ratio ($39\pm9\%$ vs. 32\% counts in the secondary) as the purported dual AGN (Section 3.1.4).

\item We analyze 340 ks of recently obtained \Chr grating data in NGC 3393 finding that the photoionized emission dominates below 3~keV even in a 4$\arcsec$ radius (Section 3.2.1). We detect a number of emission lines associated with strong photoionized emission at soft energies. We find large equivalent widths in the Fe K$\alpha$, Fe XXV $\alpha$, Fe XXVI $\alpha$ consistent with a Compton-thick AGN.

\item We use \nustar spectroscopy to study the Compton-thick nature of NGC 3393, its torus geometry, and its accretion rate (Section 3.2.2). We find both the \bntorus and \mytorus models support a Compton-thick torus seen edge-on, with a large opening angle.    Using the available \nustarsh, \Chrsh, \XMMsh, \suzakush, \bepposaxsh X-ray observations of NGC 3393, we find no evidence of variability between 1997 and 2013 (Section 3.2.3).  There is some weak evidence of variability above 10~keV between the stacked \swiftbat observations and \nustar at the 1.8$\sigma$ level dependent on the cross-calibration uncertainty.  We find a high Eddington ratio of $\approx$0.2 based on the \nev and intrinsic 2-10~keV X-ray emission (Section 3.2.4).

\end{enumerate}

	 The most important factor in the likely spurious detection of a dual AGN in NGC 3393 was the low number of X-ray counts ($<$160) at 6-7~keV and smoothing of data with a few counts per pixel on scales much smaller than the PSF (0.25$\arcsec$ vs. $0.8\arcsec$ FWHM).  Further \chandra searches for dual AGN with small angular separations ($<1\arcsec$) would benefit from studying sources with known optical or NIR counterparts such as found using high angular resolution \hst or adaptive optics imaging.  We are now extending this \chandra analysis to a larger sample of all \chandra observed \swiftbat detected AGN to test for any other spurious dual AGN detections, place upper limits on the frequency of closely separated dual AGN, and search for AGN that are significantly offset from the galaxy nucleus (Koss et al., in prep).
	
	 The combination of high-quality radio and X-ray data is useful for studying NGC 3393.  The heavily inclined torus geometry inferred from the X-ray analysis agrees well with the fact that a disk water mega-maser has been detected in NGC 3393, since masing implies a nearly edge-on view of the masing disk \citep{Kondratko:2008:87}.  At a luminosity of $\approx2\times10^{43} \ergpssh$, we would expect a small opening angle of the torus due to the obscured fraction being near its peak at this luminosity \citep[e.g.][]{Burlon:2011:58}.  It is interesting to consider whether the large opening angle of the torus found in the X-rays is linked through feedback to the strong radio jets.  Larger samples of radio bright obscured AGN will need to be studied to determine this link more thoroughly.\\
	  	 
	 \nustar has been critical for studies of local and distant AGN above 10~keV \citep[e.g.,][]{Balokovic:2014:111a,Gandhi:2014:117,Lansbury:2014:17,Puccetti:2014:26,Arevalo:2014:81,DelMoro:2014:16,Stern:2014:102}. With the full coverage of the AGN from 0.2-100~keV with deep archival \Chr and \nustar observations we can assess the importance of this type of Compton-thick  column density to different high redshift X-ray AGN surveys.  Previous studies have found that observations above 10~keV can be as successful at identifying Compton-thick AGN as deep X-ray observations of these AGN probe softer X-ray energies \citep[$<$ 10~keV, e.g.][]{Koss:2013:L26}.  The intrinsic $L_{2-10 \: \mathrm{keV}}=2.6\pm0.3\times10^{43} \ergps$ is consistent with the median luminosity in the $Chandra$ 4 Ms field at redshift z=1.6 where much of black hole growth is thought to occur.  As an unobscured source it could be detected up to redshift $z$=5.0 at SN=3 if observed near the center of the field ($<3\arcmin$) where the PSF is sharpest and a 2$\arcsec$ extraction radius encloses $>90\%$ of the flux.    We ran simulations of NGC 3393 observed on-axis with different satellites.   In the $Swift$ BAT all-sky survey, Compton-thick AGN like NGC 3393 are only detectable in the very nearby universe ($z<$0.025), whereas with $NuSTAR$ such objects are detectable out to z=0.075 with a 20 ks observation.   This simulation suggests that near the center of the deepest 4 Ms images, NGC 3393 would be detectable out to $z$=0.55 at SN=3.  However, the detection of the Fe K line at SN=3 and knowledge of the high level of obscuration and intrinsic luminosity would only be possible to redshifts $z<$0.2.  Future observations with \nustar of a large sample of nearby Compton-thick AGN ($z<$0.5) are critical to understand how large a fraction of Compton-thick AGN might be hidden in the deepest \Chr images.

\clearpage

\section*{Acknowledgments}
We thank Jim Condon, Fred Lo, and Antxon Alberdi for their useful discussion of radio data and astrometry.  We thank Harvey Tananbaum and Diab Jerius for useful discussions on the \Chr PSF and using the {\tt MARX} software.  We thank Pepi Fabbiano and Alessandro Paggi for useful discussions of previous NGC 3393 papers.   M. K. acknowledges support from the Swiss National Science Foundation (SNSF) through the Ambizione fellowship grant PZ00P2\textunderscore154799/1.  M.K. and K. S. acknowledge support from Swiss National Science Foundation (NSF) grant PP00P2 138979/1.  M.K. also acknowledges support for this work was provided by the National Aeronautics and Space Administration through \Chr Award Number AR3-14010X issued by the \Chr X-ray Observatory Center, which is operated by the Smithsonian Astrophysical Observatory for and on behalf of the National Aeronautics Space Administration under contract NAS8-03060. C.R.C. acknowledges financial support from the ALMA-CONICYT FUND Project 31100004.  We also acknowledge support from CONICYT through FONDECYT grant 3150238 (C.R.C.) and from project IC120009 "Millennium Institute of Astrophysics (MAS) funded by the Iniciativa Cient\'{\i}fica Milenio del Ministerio Econom\'{\i}a, Fomento y Turismo de Chile (C.R.C, F.E.B.).  A.C. acknowledges support from the ASI/INAF grant I/037/12/0Ð 011/13 and the Caltech Kingsley visitor program.  ST acknowledges support from the NASA postdoctoral fellowship program.  Support for the work of ET was provided by the Center of Excellence in Astrophysics and Associated Technologies (PFB 06), by the FONDECYT regular grant 1120061 and by the CONICYT Anillo project ACT1101.  M.\,B. acknowledges support from NASA Headquarters under the NASA Earth and Space Science Fellowship Program, grant NNX14AQ07H.  R.J.A. was supported by Gemini-CONICYT grant number 32120009.  This work made use of data from the \nustar mission, a project led by the California Institute of Technology, managed by the Jet Propulsion Laboratory, and funded by the National Aeronautics and Space Administration. We thank the \nustar Operations, Software and Calibration teams for support with the execution and analysis of these observations. This research has made use of the \nustar Data Analysis Software ({\tt NuSTARDAS}) jointly developed by the ASI Science DataCenter (ASDC, Italy) and the California Institute of Technology (USA). This research has made use of software provided by the \Chr X-ray Center (CXC) in the application packages {\tt CIAO}, ChIPS, and Sherpa.  The scientific results reported in this article are based on data obtained from the \Chr Data Archive.  This research made use of the XRT Data Analysis Software (XRTDAS), archival data, software and on-line services provided by the ASDC.  Based on observations obtained with XMM-Newton, an ESA science mission with instruments and contributions directly funded by ESA Member States and NASA.  The Cornell Atlas of \spitzer IRS Sources (CASSIS) is a product of the Infrared Science Center at Cornell University, supported by NASA and JPL.   We used observations made with the NASA/ESA Hubble Space Telescope, and obtained from the Hubble Legacy Archive, which is a collaboration between the Space Telescope Science Institute (STScI/NASA), the Space Telescope European Coordinating Facility (ST-ECF/ESA), and the Canadian Astronomy Data Centre (CADC/NRC/CSA).  This research made use of the \Chr Transmission Grating Catalog and archive (http://tgcat.mit.edu).  This paper is based on observations made with the Very Large Array (VLA)  and the Very Long Baseline Array (VLBA) of the National Radio Astronomy Observatory (NRAO); the NRAO is a facility of the National Science Foundation operated under cooperative agreement by Associated Universities, Inc.

{\it Facilities:}  \facility{NuSTAR}, \facility{Swift}, \facility{CXO},  \facility{HST (NICMOS, UVIS, WFC3)},  \facility{Keck:II (NIRC2)}, \facility{Spitzer}, \facility{VLBA}, \facility{VLA}, \facility{UH:2.2m}, \facility{PS1}, \facility{XMM}, \facility{Suzaku}, \facility{BeppoSAX}

\bibliographystyle{/Applications/astronat/apj/apj}
\bibliography{/Applications/astronat/bibfinal}

\end{document}